\documentclass[onecolumn,showpacs,preprintnumbers]{revtex4}
\usepackage{mathrsfs}
\usepackage{graphicx}
\usepackage{dcolumn}
\usepackage{bm}
\usepackage{epsfig}
\usepackage{amsmath,amssymb}
\usepackage{multirow}
\usepackage{placeins}
\usepackage{caption}
\usepackage{array}
\usepackage[colorlinks,
            citecolor=blue,
            anchorcolor=red,
            menucolor=red,
            linkcolor=red,
            filecolor=red,
            runcolor=red,
            urlcolor=blue,
            frenchlinks=red]{hyperref}

\newcommand{\PreserveBackslash}[1]{\let\temp=\\#1\let\\=\temp}

\newcolumntype{C}[1]{>{\PreserveBackslash\centering}p{#1}}

\newcolumntype{R}[1]{>{\PreserveBackslash\raggedleft}p{#1}}

\newcolumntype{L}[1]{>{\PreserveBackslash\raggedright}p{#1}}

\begin{document}
\title{Strong decay properties of single heavy baryons $\Lambda_{Q}$, $\Sigma_{Q}$ and $\Omega_{Q}$}
\author{Guo-Liang Yu$^{1}$}
\email{yuguoliang2011@163.com}
\author{Yan Meng$^{1}$}
\author{Zhen-Yu Li$^{2}$}
\author{Zhi-Gang Wang$^{1}$}
\email{zgwang@aliyun.com}
\author{Lu Jie$^{1}$}

\affiliation{$^1$ Department of Mathematics and Physics, North China
Electric Power University, Baoding 071003, People's Republic of
China\\$^2$ School of Physics and Electronic Science, Guizhou Education University, Guiyang 550018, People's Republic of
China}
\date{\today }

\begin{abstract}
Motivated by recent progresses in experiments in searching for the $\Omega_{c}$ baryons, we systematically analyze the strong decay behaviors of single heavy baryons $\Lambda_{Q}$, $\Sigma_{Q}$ and $\Omega_{Q}$. The two-body strong decay properties of $S$-wave, $P$-wave and some $D$-wave states are studied with the $^{3}P_{0}$ model. The results support assigning the recently observed $\Omega_{c}(3185)$ and $\Omega_{c}(3327)$ as the 2S($\frac{3}{2}^{+}$) and 1D($\frac{3}{2}^{+}$) states, respectively. In addition, the quantum numbers of many other experimentally observed baryons are also suggested according to their strong decays. Finally, some baryons which have good potentials to be observed in experiments are predicted and the possible decay channels for searching for these predicted states are also suggested.
\end{abstract}


\maketitle

\begin{Large}
\textbf{1 Introduction}
\end{Large}

Very recently, two new excited states, $\Omega_{c}(3185)$ and $\Omega_{c}(3327)$, were observed in the $\Xi^{+}_{c}K^{-}$ invariant-mass spectrum using proton-proton collision data collected by the LHCb experiment, corresponding to an integrated luminosity of 9fb$^{-1}$\cite{3185}. At the same time, another five previously observed $\Omega_{c}$ states, $\Omega_{c}(3000)$, $\Omega_{c}(3050)$, $\Omega_{c}(3065)$, $\Omega_{c}(3090)$, $\Omega_{c}(3119)$\cite{OmegaC3000} were also confirmed. Actually, scientists have made great progresses in searching for the single heavy baryons such as $\Lambda_{c}(2595)$\cite{2595}, $\Lambda_{c}(2625)$\cite{26251,26252}, $\Lambda_{c}(2765)$\cite{LambdaC2765}, $\Lambda_{c}(2940)$\cite{LambdaC29401,LambdaC29402,LambdaC29403}, $\Lambda_{b}(5912)$ and $\Lambda_{b}(5920)$\cite{59121,59122}, $\Lambda_{b}(6072)$\cite{LambdaB60721}, $\Lambda_{b}(6146)$ and $\Lambda_{b}(6152)$\cite{LambdaB6146}, $\Sigma_{c}(2800)$\cite{2800}, $\Sigma_{b}(6097)$\cite{6097}, $\Omega_{b}(6330)$, $\Omega_{b}(6316)$, $\Omega_{b}(6350)$ and $\Omega_{b}(6340)$\cite{OmegaB6316}. All of the experimentally discovered $\Lambda_{Q}$, $\Sigma_{Q}$ and $\Omega_{Q}$ baryons are collected in Tables \ref{Table I}-\ref{Table III}. The quantum numbers of some baryons have been suggested or confirmed, however the others are still unidentified.

In our previous work, we have systematically studied the mass spectra of the $\Lambda_{Q}$, $\Sigma_{Q}$ and $\Omega_{Q}$ systems with the constituent quark model\cite{GuoL}. The quantum numbers of the experimentally observed baryons were suggested and are also shown in Tables \ref{Table I}-\ref{Table III}. At the same time, some excited baryon states were also predicted in Ref.\cite{GuoL}. For example, the predicted masses are 3197 MeV for the 2$S$($\frac{3}{2}^{+}$) $\Omega_{c}$ baryon and 3304$-$3315 MeV for the 1$D$ ones. These results are consistent well with the experimental data of the newly observed $\Omega_{c}(3185)$ and $\Omega_{c}(3327)$ baryons. This implies that these two baryons can be assigned as the 2$S$($\frac{3}{2}^{+}$) and 1$D$ states, respectively. In addition, many other states which have good potentials to be found in experiments were predicted, such as the 2$P$ $\Lambda_{c}(\frac{3}{2}^{-})$, 2$P$ $\Lambda_{b}$ doublet ($\frac{1}{2}^{-}$,$\frac{3}{2}^{-}$), 1$P$ $\Sigma_{c}$/$\Sigma_{b}$, and 1$D$ $\Omega_{b}$ states\cite{GuoL}. To make a further confirmation about the assignments of the experimental states and provide more valuable infirmation for searching for these predicted baryons, it is necessary to systematically investigate the strong decay behaviors of the single heavy baryons. Fortunately, we can continue this research now with the results obtained from the quark model\cite{GuoL}.

As a phenomenological method, the $^{3}P_{0}$ quark model was developed to study the OZI-allowed hadronic decay widths\cite{3P01,3P02,3P03,3P04,3P0M01A,3P0M01B,3P0M01C,3P0M02,3P0M03}. Now, this model has been extensively used to describe the two-body strong decays of the heavy mesons in the charmonium and bottommonium systems\cite{3P0M04,3P0M05,3P0M061,3P0M062,3P0M07,3P0M081,3P0M082,3P0M09,3P0M010,3P0M011,3P0M012,3P0M013,3P0M014,3P0M015}, the baryons\cite{3P05,3P06,3P07,3P08,3P09,3P010,3P011,3P012,3P013,3P014,3P015,3P0161,3P0Z1,3P0Z2,3P0Z3,3P0Z4,3P0Z5} and even the teraquark states\cite{3P0T01}. In this work, $^{3}P_{0}$ model will be employed to study the two-body strong decay properties of the $S$-wave, $P$-wave and some $D$-wave single heavy baryons. The calculated
strong decay widths in this work will be confronted with the experimental data in the future and will be helpful in searching for new states of single heavy baryons from the Belle, BABAR, CLEO and LHCb collaborations.

The paper is organized as follows. In Section II, we give a brief review of the $^{3}P_{0}$ decay model; In Sec III, using the predicted mass spectra by quark model\cite{GuoL}, we study the two-body strong decay behaviors of the $S$-wave, $P$-wave and some $D$-wave heavy baryon states. And Sec IV is reserved for our conclusions.

\begin{Large}
\textbf{2 $^{3}P_{0}$ strong decay model}
\end{Large}

\begin{figure}[h]
\centering
\includegraphics[height=4.5cm,width=10cm]{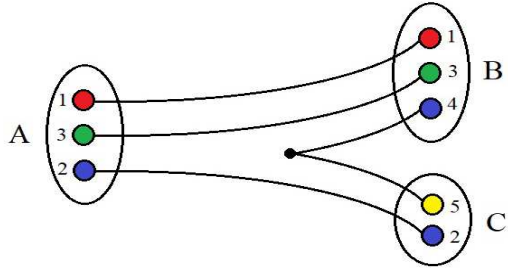}
\caption{The decay process of A$\rightarrow$B+C in the $^{3}P_{0}$ model.}
\label{Fig 1}
\end{figure}

The main idea of $^{3}P_{0}$ model model is that the strong decay takes place via the creation of a $^{3}P_{0}$ quark-antiquark pair from the vacuum. Then, this quark-antiquark pair regroups with the initial hadron A into two daughter hadrons B and C. This process is illustrated in Fig. \ref{Fig 1}.

\begin{figure}[h]
\centering
\includegraphics[height=5cm,width=6cm]{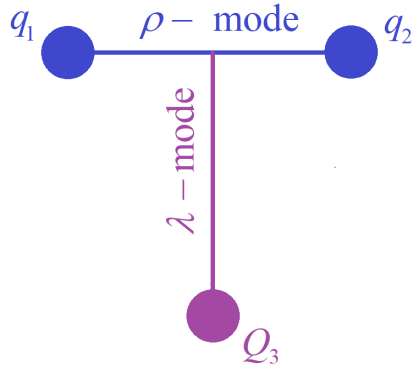}
\caption{The $\rho$- and $\lambda$-mode excitations in a single heavy baryon system. $q_{1}$ and $q_{2}$ denote the light $u$, $d$ and $s$ quarks, and $Q_{3}$ corresponds to a heavy charm or bottom quark.}
\label{SHB}
\end{figure}

In the $^{3}P_{0}$ model, the strong decay width for the process $A\rightarrow B+C$ can be written as\cite{3P05,3P09,3P010,3P011},
\begin{flalign}
\Gamma&=\pi^{2}\frac{|\emph{\textbf{p}}|}{m^{2}_{A}}\frac{1}{2J_{A}+1}\sum_{M_{J_{A}}M_{J_{B}}M_{J_{C}}}|M^{M_{J_{A}}M_{J_{B}}M_{J_{C}}}|^{2},
\end{flalign}
where $m_{A}$ and $J_{A}$ are the mass and total angular momentum of the initial baryon A, $m_{B(C)}$ and $J_{B(C)}$ are the ones for the daughter hadrons. $|\emph{\textbf{p}}|=\frac{\sqrt{[m^{2}_{A}-(m_{B}-m_{C})^{2}][m^{2}_{A}-(m_{B}+m_{C})^{2}]}}{2m_{A}}$ is the momentum of the daughter hadrons in the centre-of-mass frame. $M^{M_{J_{A}}M_{J_{B}}M_{J_{C}}}$ in Eq.(1) is the helicity amplitude, which reads\cite{3P05,3P09,3P010,3P011}
\begin{flalign}
\notag
& M^{M_{J_{A}}M_{J_{B}}M_{J_{C}}} \\ \notag
=& -F\gamma\sqrt{8E_{A}E_{B}E_{C}}\sum_{m_{l_{\rho_{A}}}}\sum_{M_{L_{A}}}\sum_{m_{l_{\rho_{B}}}}\sum_{M_{L_{B}}}\sum_{m_{s_{1}},m_{s_{3}},m_{s_{4}},m}\langle j_{A}m_{j_{A}}s_{3}m_{s_{3}}|J_{A}M_{J_{A}}\rangle  \\ \notag
&\times\langle l_{\rho_{A}}m_{l_{\rho_{A}}}l_{\lambda_{A}}m_{l_{\lambda_{A}}}|L_{A}M_{L_{A}}\rangle\langle L_{A}M_{L_{A}}S_{12}M_{S_{12}}| j_{A}m_{j_{A}}\rangle\langle s_{1}m_{s_{1}}s_{2}m_{s_{2}}| S_{12}M_{S_{12}}\rangle \\ \notag
&\times\langle j_{B}m_{j_{B}}m_{3}m_{m_{3}}|J_{B}M_{J_{B}}\rangle\langle l_{\rho_{B}}m_{l_{\rho_{B}}}l_{\lambda_{B}}m_{l_{\lambda_{B}}}|L_{B}M_{L_{B}}\rangle\langle L_{B}M_{L_{B}}S_{14}M_{S_{14}}| j_{B}m_{j_{B}}\rangle  \\ \notag
&\times\langle s_{1}m_{s_{1}}s_{4}m_{s_{4}}| S_{14}M_{S_{14}}\rangle\langle 1m;1-m|00\rangle\langle s_{4}m_{s_{4}}s_{5}m_{s_{5}}|1-m\rangle \\ \notag
&\times\langle L_{C}M_{L_{C}}S_{C}M_{S_{C}}|J_{C}M_{J_{C}}\rangle\langle s_{2}m_{s_{2}}s_{5}m_{s_{5}}|S_{C}M_{S_{C}}\rangle \times \\
&\times\langle \varphi^{1,4,3}_{B}\varphi^{2,5}_{C}|\varphi^{1,2,3}_{A}\varphi^{4,5}_{0}\rangle \times I_{M_{L_{B}},M_{L_{C}}}^{M_{L_{A}},m}(\emph{\textbf{p}}).
\end{flalign}
In the above equation, $\emph{\textbf{s}}_{i}$ is the spin of the $i$th quark, $\emph{\textbf{l}}_{\rho}$ and $\emph{\textbf{l}}_{\lambda}$ denote the orbital angular momentum between two light quarks, and between the heavy quark and light quark subsystem(Fig. \ref{SHB}), respectively. $\emph{\textbf{L}}$ is the total orbital angular momentum, $\emph{\textbf{j}}$ represents total angular momentum of $\emph{\textbf{L}}$ and the total spin $\emph{\textbf{S}}_{ij}$ of the light quark subsystem, $\emph{\textbf{J}}$ is the total angular momentum of a baryon. The Clebsch-Gordan coefficients in Eq.(2) indicate the conservation of the angular momentum, \emph{e}.\emph{g}., $\emph{\textbf{s}}_{1}+\emph{\textbf{s}}_{2}=\emph{\textbf{S}}_{12}$, $\emph{\textbf{l}}_{\rho A}+\emph{\textbf{l}}_{\lambda A}=\emph{\textbf{L}}_{A}$, $\emph{\textbf{S}}_{12}+\emph{\textbf{L}}_{A}=\emph{\textbf{j}}_{A}$ and $\emph{\textbf{s}}_{3}+\emph{\textbf{j}}_{A}=\emph{\textbf{J}}_{A}$. $F$ is a factor equal to $2$ when each one of the two quarks in C has isospin $\frac{1}{2}$ , and $F=1$ when one of the two quarks in C has isospin $0$\cite{3P04,3P010}. $\langle \varphi^{1,4,3}_{B}\varphi^{2,5}_{C}|\varphi^{1,2,3}_{A}\varphi^{4,5}_{0}\rangle$ is the flavor matrix element of the flavor wave functions $\varphi_{i}$($i$=$A$,$B$,$C$,0), which has the following relation with the isospin matrix element $\langle I_{B}I_{B}^{3} I_{C}I_{C}^{3}|I_{A}I_{A}^{3}\rangle$\cite{3P04,3P09},
\begin{flalign}
& \langle \varphi^{1,4,3}_{B}\varphi^{2,5}_{C}|\varphi^{1,2,3}_{A}\varphi^{4,5}_{0}\rangle=\mathcal{F}^{(I^{A},I^{B},I^{C})}\langle I_{B}I_{B}^{3} I_{C}I_{C}^{3}|I_{A}I_{A}^{3}\rangle,
\end{flalign}
where
\begin{flalign}
\mathcal{F}^{(I^{A},I^{B},I^{C})}=\emph{f}\times(-1)^{I_{13}+I_{C}+I_{A}+I_{2}} \times [\frac{1}{2}(2I_{C}+1)(2I_{B}+1)]^{1/2} \times \left \{ \begin{array}{ccc} I_{13} & I_{B} &I_{4}\\ I_{C} & I_{2} &I_{A}  \end{array}\right \}.
\end{flalign}
Here, $I_{A}$, $I_{B}$ and $I_{C}$ represent the isospins of the initial baryon, the final baryon and the final meson. $I_{13}$, $I_{2}$ and $I_{4}$ denote
the isospins of relevant quarks, respectively. $f$ takes the value $f=\big(\frac{2}{3}\big)^{1/2}$ if the isospin of the created quark is $\frac{1}{2}$, and $f=-\big(\frac{1}{3}\big)^{1/2}$ for the isospin of $0$\cite{3P04,3P010}. In Eq.(2), the space integral is written as\cite{3P05,3P09,3P010,3P011},
\begin{flalign}
\notag
I_{M_{L_{B}},M_{L_{C}}}^{M_{L_{A}},m}(\emph{\textbf{p}})&=\int d\emph{\textbf{p}}_{1}d\emph{\textbf{p}}_{2}d\emph{\textbf{p}}_{3}d\emph{\textbf{p}}_{4}d\emph{\textbf{p}}_{5} \delta^{3}(\emph{\textbf{p}}_{1}+\emph{\textbf{p}}_{2}+\emph{\textbf{p}}_{3}-\emph{\textbf{p}}_{A})\delta^{3}(\emph{\textbf{p}}_{4}+\emph{\textbf{p}}_{5})   \\ \notag
&\times \delta^{3}(\emph{\textbf{p}}_{1}+\emph{\textbf{p}}_{4}+\emph{\textbf{p}}_{3}-\emph{\textbf{p}}_{B})\delta^{3}(\emph{\textbf{p}}_{2}+\emph{\textbf{p}}_{5}-\emph{\textbf{p}}_{C})\Psi^{*}_{B}(\emph{\textbf{p}}_{1},\emph{\textbf{p}}_{4},\emph{\textbf{p}}_{3})\Psi^{*}_{C}(\emph{\textbf{p}}_{2},\emph{\textbf{p}}_{5})   \\
&\times \Psi_{A}(\emph{\textbf{p}}_{1},\emph{\textbf{p}}_{2},\emph{\textbf{p}}_{3})y_{lm}(\frac{\emph{\textbf{p}}_{4}-\emph{\textbf{p}}_{5}}{2}).
\end{flalign}
where $\emph{\textbf{p}}_{i}$ represents momentum of the $i$th quark and $\emph{\textbf{p}}_{A}$, $\emph{\textbf{p}}_{B}$ and $\emph{\textbf{p}}_{C}$ are the momentum of hadrons. In this work, the simple harmonic oscillator(SHO) wave function is chosen as the spatial part of the baryons\cite{Bpara},
\begin{flalign}
&\Psi(\emph{\textbf{p}})=N\Psi_{n_{\rho}l_{\rho}m_{l_{\rho}}}(\emph{\textbf{p}}_{\rho})\Psi_{n_{\lambda}l_{\lambda}m_{l_{\lambda}}}(\emph{\textbf{p}}_{\lambda}).
\end{flalign}
Here, $N$ is a normalization coefficient of the wave function, $\emph{\textbf{p}}_{\rho}$ and $\emph{\textbf{p}}_{\lambda}$ represent the relative momentum between two light quarks, and between the heavy quark and the center of mass of two light quarks(Fig. \ref{SHB}), respectively. The relative wave function in the above equation is written as,
\begin{flalign}
&\Psi_{nlm_{l}}(\emph{\textbf{p}})=(-1)^{n}(-i)^{l}R^{l+\frac{3}{2}}\sqrt{\frac{2n!}{\Gamma(n+l+\frac{3}{2})}}\mathrm{exp}(-\frac{R^{2}\emph{\textbf{p}}^{2}}{2})\times L_{n}^{l+1/2}(R^{2}\emph{\textbf{p}}^{2})|\emph{\textbf{p}}|^{l}Y_{lm_{l}}(\Omega_{p}), \\ \notag
\end{flalign}
where $L_{n}^{l+1/2}(R^{2}\emph{\textbf{p}}^{2})$ is the Laguerre polynomial function, and $Y_{lm_{l}}(\Omega_{p})$ represents the spherical harmonic function. The relation between the solid harmonica polynomial $y_{lm}(\emph{\textbf{p}})$ and $Y_{lm_{l}}(\Omega_{p})$ can be written as $y_{lm}(\emph{\textbf{p}})=|\emph{\textbf{p}}|^{l}Y_{lm_{l}}(\Omega_{p})$.

In the heavy quark limit, the heavy quark in a single heavy baryon is decoupled from two light quarks. Under this picture, the dynamics of a single heavy baryon are commonly separated into two parts, which is illustrated in Fig. \ref{SHB}. The degree of freedom between two light quarks($q_{1}$ and $q_{2}$) is called $\rho$-mode, while the degree of freedom between the center of mass of two light quarks and the heavy quark is called $\lambda$-mode. For $P$-wave baryons, there are two orbital excitation modes $\lambda$- and $\rho$-mode with ($l_{\rho}$,$l_{\lambda}$)=(0,1) and (1,0) respectively. While there are three excitation modes for $D$-wave baryons with ($l_{\rho}$,$l_{\lambda}$)=(0,2), (2,0) and (1,1), which are called the $\lambda$-mode, $\rho$-mode and $\lambda$-$\rho$ mixing mode, respectively. It is indicated that the lowest state of a single heavy baryon is dominated by the $\lambda$-mode excitations and almost all experimentally observed single heavy baryons can be interpreted as $\lambda$-mode excited states\cite{GuoL,3P013}. Thus, only the strong decays of single heavy baryons with $\lambda$-mode orbital excitations are considered in this work.

\begin{Large}
\textbf{3 Numerical results and discussions}
\end{Large}

The final results of $^{3}P_{0}$ model depend on some input parameters such as the quark pair ($q\overline{q}$) creation strength $\gamma$, the SHO wave function scale parameter $R$, and the masses of the hadrons. As for the quark pair creation strength, we take the universal value $\gamma=13.4$\cite{3P0M01A,3P05,3P010}. For the heavy baryons, the parameters $R_{\lambda,\rho}$ can be fixed to reproduce the mass splitting through the contact term in the potential model\cite{Bpara}. Their values are taken as $R_{\lambda,\rho}=1.67$ GeV$^{-1}$ for $1S$-wave baryon, $R_{\lambda,\rho}=2.0$ GeV$^{-1}$ for $P$-wave baryon and $R_{\lambda,\rho}=2.5$ GeV$^{-1}$ for $2S$- and $2D$-wave baryons\cite{3P05,3P010,Bpara}. As for the light mesons, the value of $R$ was suggested to be $2.5$ GeV$^{-1}$ in Ref.\cite{3P0M01A,3P0M01B,3P0M01C}, where it was determined by performing a series of least squares fits
of the model predictions to the decay widths of 28 of the best known meson decays. For heavier mesons, their values are different from light meson's and are taken as $R_{B}=1.59$ GeV$^{-1}$, $R_{B_{s}}=1.53$ GeV$^{-1}$, $R_{D}=1.67$ GeV$^{-1}$ and $R_{D_{s}}=1.63$ GeV$^{-1}$\cite{3P017,3P018}. All these $R$ values of heavy mesons were predicted by the relativistic quark model\cite{3P017,3P018}. With the predicted masses in Refs.\cite{GuoL}, we study the OZI-allowed strong decay behaviors of the $1S$, $2S$, $1P$, $2P$ states and some of the $1D$, $2D$ states. The results are presented in Tables \ref{Table IV}-\ref{Table XXI} of Appendix A. In these tables, the single heavy baryons are denoted as $\Lambda_{Qj}^{L\lambda^{\prime}(\rho^{\prime})}$, $\Sigma_{Qj}^{L\lambda^{\prime}(\rho^{\prime})}$ and $\Omega_{Qj}^{L\lambda^{\prime}(\rho^{\prime})}$, where the superscript $L$ is the total orbital angular momentum with $L=l_{\rho}+l_{\lambda}$. The superscripts $\lambda^{\prime}$ and $\rho^{\prime}$ denote the first radial excitation with ($n_{\rho}$,$n_{\lambda}$)=($0$, $1$) and ($1$, $0$), respectively. The subscript $j$ represents the total angular momentum of light quarks which satisfies $j=L+S$. The experimental information\cite{article2B,article2A} and predictions from quark model\cite{GuoL} for $\Lambda_{Q}$, $\Sigma_{Q}$ and $\Omega_{Q}$ baryons are collected in \ref{Table I}-\ref{Table III}, and the predicted widths at present work by $^{3}P_{0}$ model are also shown in the last column in these tables.

\begin{large}
\textbf{3.1 $\Lambda_{Q}$ states}
\end{large}

For $\Lambda_{Q}$ systems, we can see from Tables \ref{Table I}, \ref{Table IV}-\ref{Table VII} that the predicted decay widths are roughly compatible with the experimental data. For $\Lambda_{c}(2765)$, it was suggested as a $2S$($\frac{1}{2}^{+}$) state by quark model\cite{GuoL}. The predicted total width of the $2S$-wave $\Lambda_{c0}^{\lambda^{\prime}}$($\frac{1}{2}^{+}$) state is $25.34$ MeV(Table \ref{Table IV}), which is lower than the experimental data. Considering the theoretical uncertainty, it is reasonable to assign $\Lambda_{c}(2765)$ as the $2S$($\frac{1}{2}^{+}$). The theoretical total width for $2P$-wave $\Lambda_{c1}^{\lambda^{\prime}}$($\frac{1}{2}^{-}$) is $12.40$ MeV(Table \ref{Table IV}), which is compatible with the experimental data of $\Lambda_{c}(2940)$. Thus, $\Lambda_{c}(2940)$ can be assigned as the $2P$($\frac{1}{2}^{-}$) state.  As for the $\Lambda_{c}(2860)$ and $\Lambda_{c}(2880)$, they were interpreted as a $1D$ doublet ($\frac{3}{2}^{+}$,$\frac{5}{2}^{+}$) by other collaborations\cite{3P010}. Our predicted mass spectrum in quark model also supports this conclusion\cite{GuoL}. Under this assignment, the theoretical total width for $\Lambda_{c}(2860)$ is $9.03$ MeV(Tables \ref{Table I} and \ref{Table V}) which is consistent with the results of other collaborations\cite{3P010}. However, this value is much lower than the experimental data which is about $67$ MeV. We hope this divergence can be clarified in the future if more theoretical and experimental efforts were devoted into this problem.
\begin{table*}[h]
\begin{ruledtabular}\caption{Experimental information of $\Lambda_{Q}$(Q=c,b) baryons, and the predictions by quark model and $^{3}P_{0}$ model. All values are in units of MeV}
\label{Table I}
\begin{tabular}{c c c c c |c c |c}
\multicolumn{5}{c}{Experimental information\cite{article2B,article2A}} & \multicolumn{2}{c}{Quark model\cite{GuoL}} &$^{3}P_{0}$ model \\
States  & $J^{P}$ &Mass  &Width &Decay channels  &  $J^{P}$    &  Mass  & Width\\ \hline
$\Lambda_{c}^{+}$&$\frac{1}{2}^{+}$&2286.46$\pm$0.14&/&weak &$\frac{1}{2}^{+}$(1S) & 2288& -\\
$\Lambda_{c}(2595)^{+}$&$\frac{1}{2}^{-}$&2592.25$\pm$0.28&2.59$\pm$0.30$\pm$0.47&$\Sigma_{c}^{++,0}$$\pi^{-,+}$ &$\frac{1}{2}^{-}$(1P)&2596& 11.44\\
$\Lambda_{c}(2625)^{+}$&$\frac{3}{2}^{-}$&2628.11$\pm$0.19&$<0.97$&$\Sigma_{c}^{++,0}$$\pi^{-,+}$ &$\frac{3}{2}^{-}$(1P)&2631&1.0$\times10^{-3}$\\
$\Lambda_{c}(2765)^{+}$&$?^{?}$&2766.6$\pm$2.4&50& $\Sigma^{++/0}_{c}\pi^{\mp}$ &$\frac{1}{2}^{+}$(2S)& 2764&25.34\\
$\Lambda_{c}(2860)^{+}$&$\frac{3}{2}^{+}$&$2856.1^{+2.0}_{-1.7}$$\pm$$0.5^{+1.1}_{-5.6}$&$67.6^{+10.1}_{-8.1}$$\pm$$1.4^{+5.9}_{-20.0}$&$D^{0}$p &$\frac{3}{2}^{+}$(1D)&2875&9.03\\
$\Lambda_{c}(2880)^{+}$&$\frac{5}{2}^{+}$&2881.63$\pm$0.24&$5.6^{+0.8}_{-0.6}$&$\Sigma_{c}^{(*)++,0}$$\pi^{-,+}$,$D^{0}$p &$\frac{5}{2}^{+}$(1D)&2891&7.22\\
$\Lambda_{c}(2940)^{+}$&$?^{?}$&$2939.6^{+1.3}_{-1.5}$&$20^{+6}_{-5}$&$\Sigma_{c}^{++,0}$$\pi^{-,+}$  &$\frac{1}{2}^{-}$(2P)&2988& 12.40\\
$\Lambda_{b}(6072)^{0}$&$?^{?}$&6072.3$\pm$2.9&72$\pm$11&$\Lambda_{b}^{0}\pi^{+}\pi^{-}$  &$\frac{1}{2}^{+}$(2S)&6041& 8.82\\
$\Lambda_{b}(6146)^{0}$&$\frac{3}{2}^{+}$&6146.17$\pm$0.4&2.9$\pm$1.3&$\Lambda_{b}^{0}\pi^{+}\pi^{-}$ &$\frac{3}{2}^{+}$(1D)&6137& 5.99\\
$\Lambda_{b}(6152)^{0}$&$\frac{5}{2}^{+}$&6152.5$\pm$0.4&2.1$\pm$0.9&$\Lambda_{b}^{0}\pi^{+}\pi^{-}$ &$\frac{5}{2}^{+}$(1D)&6145& 5.43\\
\end{tabular}
\end{ruledtabular}
\end{table*}

The bottom baryons $\Lambda_{b}$(6072), $\Lambda_{b}$(6146) and $\Lambda_{b}$(6152) were observed in the $\Lambda_{b}^{0}\pi^{+}\pi^{-}$  invariant mass spectrum. It is known that this three-body decay to $\Lambda_{b}^{0}\pi^{+}\pi^{-}$ can take place according to intermediate $\Sigma_{b}^{\pm}$ and $\Sigma_{b}^{*\pm}$ states. For $\Lambda_{b}(6152)$ as an example, the $\Lambda_{b}(6152)\rightarrow \Sigma_{b}\pi$ and $\Lambda_{b}(6152)\rightarrow \Sigma_{b}^{*}\pi$ processes have been clearly visible\cite{LambdaB6146}. Thus, the two-body strong decay properties are commonly estimated by the $^{3}P_{0}$ model, which serves as an important information to understand the nature of these newly observed baryons\cite{3P011,3P013,3P015,3P0Z2,LambdaB60722}. In many references, $\Lambda_{b}(6146)$ and $\Lambda_{b}(6152)$ were suggested to be the $1D$-wave doublet ($\frac{3}{2}^{+}$,$\frac{5}{2}^{+}$) which are the partners of the $\Lambda_{c}(2860)$ and $\Lambda_{c}(2880)$\cite{3P013,3P015}. In this work, the predicted total widths for this doublet are $5.99$ MeV and $5.43$ MeV(Tables \ref{Table I} and \ref{Table VII}), respectively. These values are comparable with the experimental measurements and consistent well with the predictions of other collaborations\cite{3P013,3P015}. However, the LHCb announced that they did not observe significant $\Lambda_{b}(6146)\rightarrow \Sigma_{b}^{\pm}\pi^{\pm}$ signals in their experiments. This divergence between experiments and model prediction needs further confirmation.

As for $\Lambda_{b}(6072)$, the LHCb Collaboration suggested that it can be assigned as the first radial excitation of $\Lambda_{b}$ baryon\cite{LambdaB60721}, 2S($\frac{1}{2}^{+}$) state. The predicted masses for the 2S($\frac{1}{2}^{+}$), 1P($\frac{1}{2}^{-}$) and 1P($\frac{3}{2}^{-}$) $\Lambda_{b}$ baryons by constituent quark model\cite{GuoL} are 6041 MeV, 5898 MeV and 5913 MeV, respectively. The measured mass of $\Lambda_{b}(6072)$ is 6072 MeV\cite{LambdaB60721}, which is compatible with the prediction for 2S($\frac{1}{2}^{+}$) state from quark model. However, the theoretical width for 2S($\frac{1}{2}^{+}$) state is 8.82 MeV in this work(Tables \ref{Table I} and \ref{Table VI}), which is much smaller than the experimental data. In Ref.\cite{LambdaB60722}, the predicted width for 2S($\frac{1}{2}^{+}$) state was 9.27 MeV, which is also significantly smaller than experimental data and consistent with our results. In their studies, they also treated $\Lambda_{b}(6072)$ as a 1P-wave state and predicted its total width to be 72 MeV by considering the mixing mechanism. However, the theoretical masses of 1P-wave $\Lambda_{b}$ baryons\cite{GuoL} are not consistent with experiments. This divergence between theoretical prediction and experimental data suggests that the nature of $\Lambda_{b}(6072)$ needs further verification by different theoretical methods and experiments.

It is also shown in Tables \ref{Table IV}-\ref{Table VI}, the predicted total widths for $2P$-wave states $\Lambda_{c1}^{1\lambda^{\prime} }$($\frac{3}{2}^{-}$), $\Lambda_{b1}^{1\lambda^{\prime} }$($\frac{1}{2}^{-}$) and $\Lambda_{b1}^{1\lambda^{\prime} }$($\frac{3}{2}^{-}$) are $13.31$ MeV, $5.97$ MeV and $5.81$ MeV, respectively. These widths are relatively narrow, which indicates these states have good potentials to be discovered in the future. To be more specific, the main decay channels are $\Sigma_{c}^{*+,0}\pi^{0,+}$, $D^{*0}p$ and $D^{*+}n$ for $\Lambda_{c1}^{1\lambda^{\prime} }$($\frac{3}{2}^{-}$), while $\Sigma_{b}^{+,0,-}\pi^{-,0,+}$ and $\Sigma_{b}^{*+,0,-}\pi^{-,0,+}$ are the main decay modes for $\Lambda_{b1}^{1\lambda^{\prime} }$($\frac{1}{2}^{-}$) and $\Lambda_{b1}^{1\lambda^{\prime} }$($\frac{3}{2}^{-}$), respectively.

\begin{large}
\textbf{3.2 $\Sigma_{Q}$ states}
\end{large}
\begin{table*}[h]
\begin{ruledtabular}\caption{Experimental information of $\Sigma_{Q}$(Q=c,b) baryons, and the predictions by quark model and $^{3}P_{0}$ model. All values are in units of MeV}
\label{Table II}
\begin{tabular}{c c c c c |c c |c}
\multicolumn{5}{c}{Experimental information\cite{article2B,article2A}} & \multicolumn{2}{c}{Quark model\cite{GuoL}} &$^{3}P_{0}$ model \\
States  & $J^{P}$ &Mass  &Width &Decay channels  &  $J^{P}$    &  Mass  & width\\ \hline
$\Sigma_{c}(2455)^{++}$&$\frac{1}{2}^{+}$&2453.97$\pm$0.14&$1.89^{+0.09}_{-0.18}$&$\Lambda_{c}^{+}$$\pi$ &\multirow{3}{*}{$\frac{1}{2}^{+}$(1S)}&\multirow{3}{*}{2457}& \multirow{3}{*}{-}\\
$\Sigma_{c}(2455)^{+}$&$\frac{1}{2}^{+}$&2452.9$\pm$0.4&$<$4.6&$\Lambda_{c}^{+}$$\pi$ &\multirow{3}{*}{}& \\
$\Sigma_{c}(2455)^{0}$&$\frac{1}{2}^{+}$&2453.75$\pm$0.14&$1.83^{+0.11}_{-0.19}$&$\Lambda_{c}^{+}$$\pi$&\multirow{3}{*}{}&\\ \hline
$\Sigma_{c}(2520)^{++}$&$\frac{3}{2}^{+}$&$2518.41^{+0.21}_{-0.19}$&$14.78^{+0.30}_{-0.40}$&$\Lambda_{c}^{+}$$\pi$&\multirow{3}{*}{$\frac{3}{2}^{+}(1S)$}&\multirow{3}{*}{2532}& \multirow{3}{*}{12.90}\\
$\Sigma_{c}(2520)^{+}$&$\frac{3}{2}^{+}$&2517.5+2.3&$<$17&$\Lambda_{c}^{+}$$\pi$&\multirow{3}{*}{}&\\
$\Sigma_{c}(2520)^{0}$&$\frac{3}{2}^{+}$&2518.48$\pm$0.20&$15.3^{+0.4}_{-0.5}$&$\Lambda_{c}^{+}$$\pi$&\multirow{3}{*}{}&\\ \hline
$\Sigma_{c}(2800)^{++}$&$?^{?}$&$2801^{+4}_{-6}$&$75^{+22}_{-17}$&$\Lambda_{c}^{+}$$\pi$&\multirow{3}{*}{$\frac{3}{2}^{-}(1P)$}&\multirow{3}{*}{2802}&\multirow{3}{*}{60$\sim$70} \\
$\Sigma_{c}(2800)^{+}$&$?^{?}$&$2792^{+14}_{-5}$&$62^{+60}_{-40}$&$\Lambda_{c}^{+}$$\pi$&\multirow{3}{*}{}&\\
$\Sigma_{c}(2800)^{0}$&$?^{?}$&$2806^{+5}_{-7}$&$72^{+22}_{-15}$&$\Lambda_{c}^{+}$$\pi$&\multirow{3}{*}{}&\\ \hline
$\Sigma_{b}^{-}$&$\frac{1}{2}^{+}$&5815.64$\pm$0.27&5.3$\pm$0.5&$\Lambda_{b}^{0}\pi$ &\multirow{2}{*}{$\frac{1}{2}^{+}$(1S)}&\multirow{2}{*}{5820}& \multirow{2}{*}{-}\\
$\Sigma_{b}^{+}$&$\frac{1}{2}^{+}$&5810.56$\pm$0.25&5.0$\pm$0.5&$\Lambda_{b}^{0}\pi$ &\multirow{2}{*}{}&\\ \hline
$\Sigma_{b}^{*+}$&$\frac{3}{2}^{+}$&5830.32$\pm$0.27&9.4$\pm$0.5&$\Lambda_{b}^{0}\pi$ &\multirow{2}{*}{$\frac{3}{2}^{+}$(1S)}&\multirow{2}{*}{5849}& \multirow{2}{*}{14.00}\\
$\Sigma_{b}^{*-}$&$\frac{3}{2}^{+}$&5834.74$\pm$0.30&10.4$\pm$0.8&$\Lambda_{b}^{0}\pi$ &\multirow{2}{*}{}&\\ \hline
$\Sigma_{b}(6097)^{-}$&$?^{?}$&6098.0$\pm$1.8&29$\pm$4&$\Lambda_{b}\pi,\Sigma_{b}\pi,\Sigma_{b}^{*}\pi$ &\multirow{2}{*}{$\frac{3}{2}^{-}$(1P)}&\multirow{2}{*}{6104} & \multirow{2}{*}{30}\\
$\Sigma_{b}(6097)^{+}$&$?^{?}$&6095.8$\pm$1.7&31$\pm$6&$\Lambda_{b}\pi,\Sigma_{b}\pi,\Sigma_{b}^{*}\pi$ \\
\end{tabular}
\end{ruledtabular}
\end{table*}

As for the $\Sigma_{Q}$ baryons, the lowest $S$-wave states $\frac{1}{2}^{+}$ and $\frac{3}{2}^{+}$ have been observed and confirmed\cite{article2A,article2B}. However, the spin-parities of experimentally observed $\Sigma_{c}(2800)$ and $\Sigma_{b}(6097)$ need confirmation in more ways\cite{3P09,3P012}. According to the predictions by quark model, both $\Sigma_{c}(2800)$ and $\Sigma_{b}(6097)$ can be accommodated in the mass spectra as lowest-lying $P$-wave states. Yet it is noted that there are five $1P$-wave states in quark model, where their masses are close to each other\cite{GuoL}. The $\Sigma_{b}(6097)$ is proposed to be a $1P$($\frac{3}{2}^{-}$)$_{j=2}$ state by quark model and so does $\Sigma_{c}(2800)$. From Tables \ref{Table VIII}-\ref{Table XIII}, we can see that both  $\Sigma_{c}(2800)$ and $\Sigma_{b}(6097)$ are impossible the spin singlet $J^{P}=\frac{1}{2}^{-}$ or the spin doublet ($\frac{1}{2}^{-}$,$\frac{3}{2}^{-}$)$_{j=1}$ because of their large theoretical widths. For $\Sigma_{b}$ baryons, the predicted total widths for $\Sigma_{b2}^{1}$($\frac{3}{2}^{-}$) or $\Sigma_{b2}^{1}$($\frac{5}{2}^{-}$) are $15.64$ MeV and $18.65$ MeV respectively(Table \ref{Table XIII}), which are close to each other and at the same order of magnitude with the experimental data. Because the predicted mass for $\Sigma_{b2}^{1}$($\frac{3}{2}^{-}$) by quark model is closer to the measured results\cite{GuoL}, the $\Sigma_{b2}^{1}$($\frac{3}{2}^{-}$) is a better candidate for $\Sigma_{b}(6097)$. As for $\Sigma_{c}(2800)$, its situation is very similar with that of $\Sigma_{b}(6097)$\cite{28001,28002,28003} and the possible assignment for it is $\Sigma_{c2}^{1}$($\frac{3}{2}^{-}$). If these assignments for $\Sigma_{c}(2800)$ and $\Sigma_{b}(6097)$ are true, their predicted total widths are still lower than measured values. Especially for $\Sigma_{c}(2800)$, its theoretical width is $19.13$ MeV, which is much lower than excremental data. The first interpretation of this deviation is the uncertainties of the $^{3}P_{0}$ model. In the following, we will see that the result from the $^{3}P_{0}$ model may be a factor of $2\sim3$ off the experimental width. Another interpretation of this problem is the mixing mechanism of the quark model states. We know that the physical resonances can be the mixing of the quark model states with the same $J^{P}$\cite{WL},
\begin{equation}
\left(
\begin{array}{c}
|1P \frac{1}{2}^{-}\rangle_{1} \\
|1P \frac{1}{2}^{-}\rangle_{2}
\end{array}
\right)=\left(
\begin{array}{c c}
\mathbf{cos}\theta \quad \mathbf{sin}\theta \\
-\mathbf{sin}\theta \quad \mathbf{cos}\theta
\end{array}
\right)\left(
\begin{array}{c}
|\frac{1}{2}^{-},j=0\rangle \\
|\frac{1}{2}^{-},j=1\rangle
\end{array}
\right),
\end{equation}
\begin{equation}
\left(
\begin{array}{c}
|1P \frac{3}{2}^{-}\rangle_{1} \\
|1P \frac{3}{2}^{-}\rangle_{2}
\end{array}
\right)=\left(
\begin{array}{c c}
\mathbf{cos}\theta \quad \mathbf{sin}\theta \\
-\mathbf{sin}\theta \quad \mathbf{cos}\theta
\end{array}
\right)\left(
\begin{array}{c}
|\frac{3}{2}^{-},j=1\rangle \\
|\frac{3}{2}^{-},j=2\rangle
\end{array}
\right).
\end{equation}
\begin{figure}[h]
\begin{minipage}[t]{0.45\linewidth}
\centering
\includegraphics[height=5cm,width=7cm]{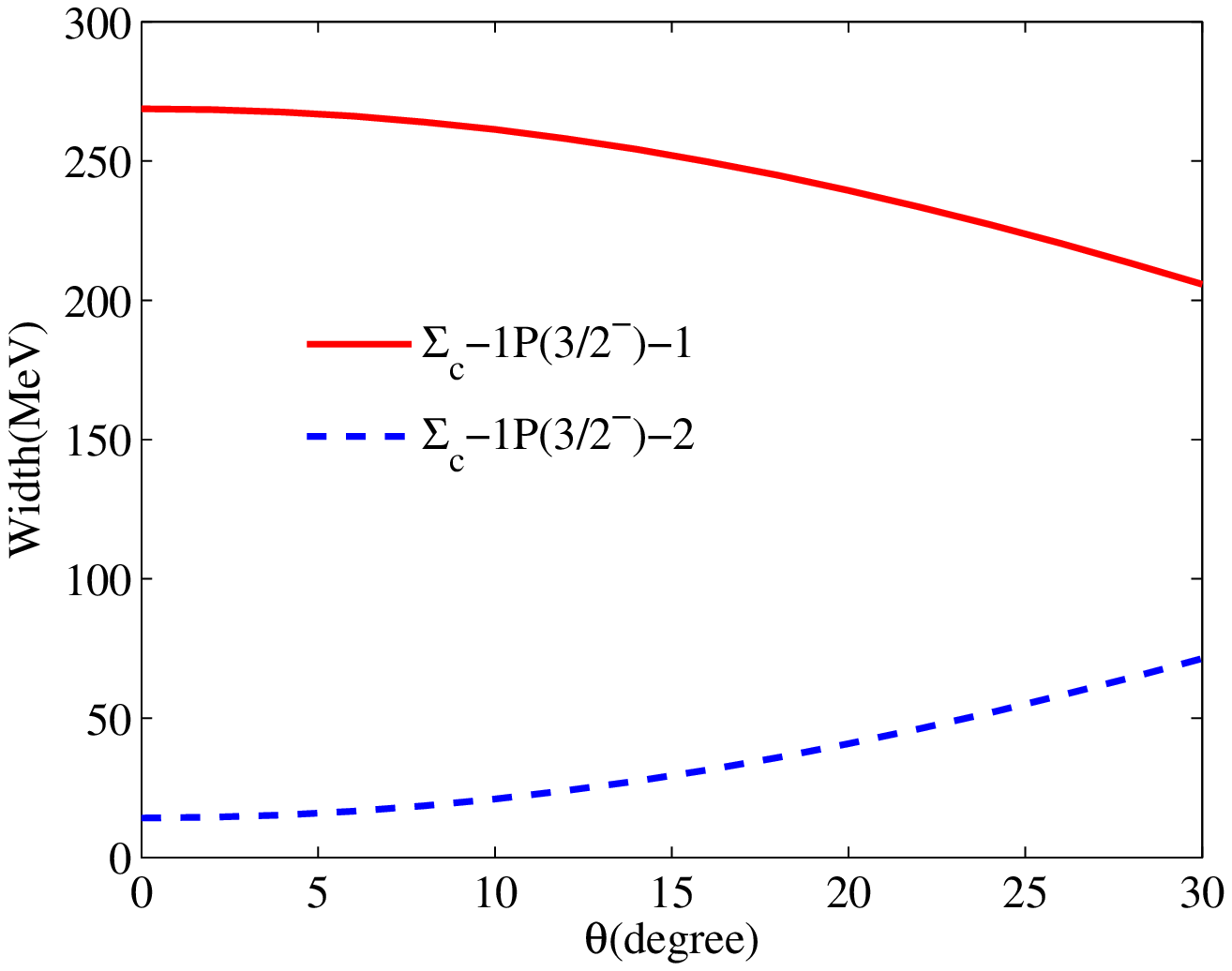}
\caption{The total decay widths of $|1P\frac{3}{2}^{-}\rangle_{1}$ and $|1P\frac{3}{2}^{-}\rangle_{2}$ for $\Sigma_{c}$ as functions
of the mixing angle $\theta$ in the range $0^{\circ}\sim30^{\circ}$.}
\label{Fig 2}
\end{minipage}
\hfill
\begin{minipage}[t]{0.45\linewidth}
\centering
\includegraphics[height=5cm,width=7cm]{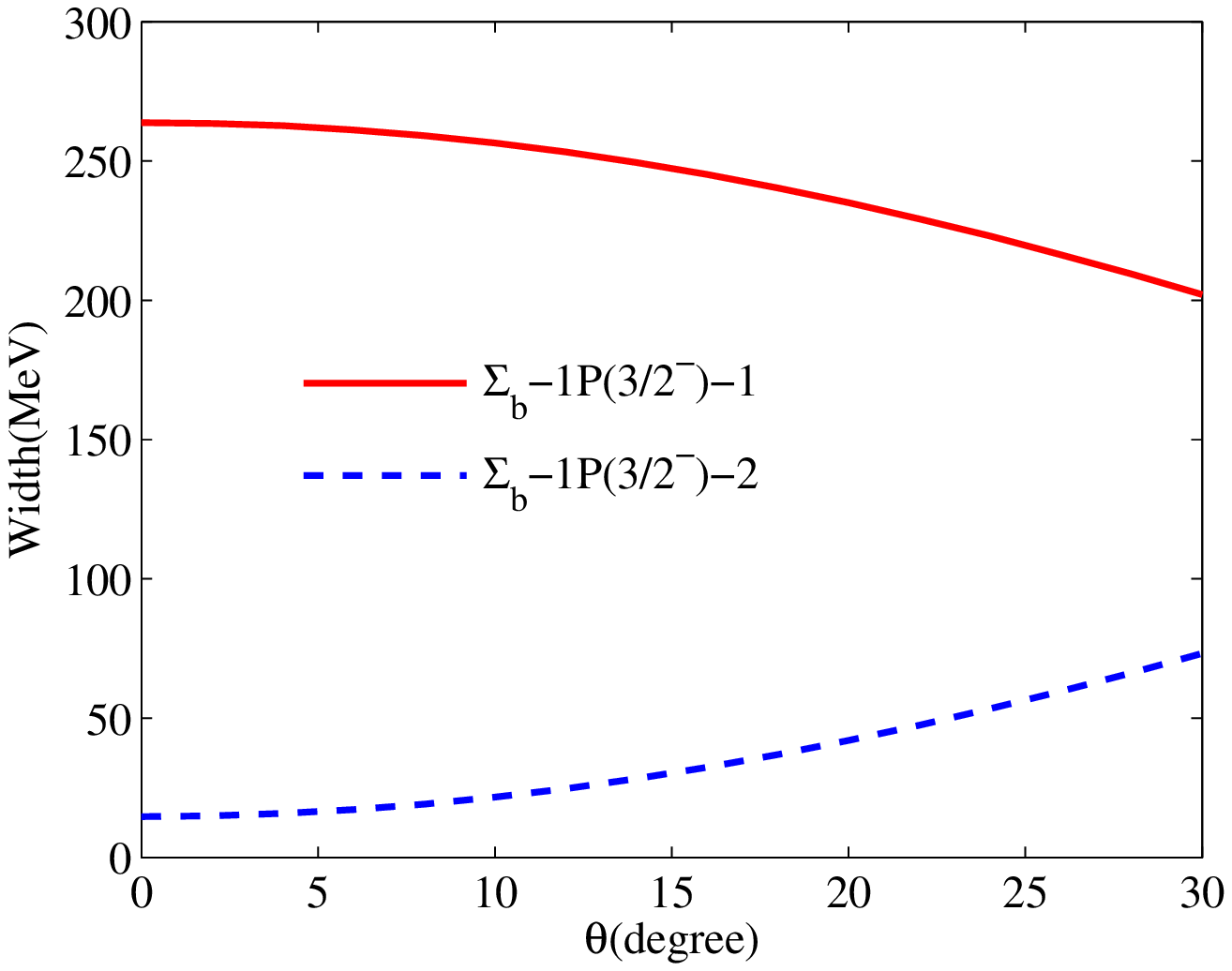}
\caption{The total decay widths of $|1P\frac{3}{2}^{-}\rangle_{1}$ and $|1P\frac{3}{2}^{-}\rangle_{2}$ for $\Sigma_{b}$ as functions
of the mixing angle $\theta$ in the range $0^{\circ}\sim30^{\circ}$.}
\label{Fig 3}
\end{minipage}
\end{figure}
That is to say, state mixing can occur between $|\frac{3}{2}^{-},j=1\rangle$ and $|\frac{3}{2}^{-},j=2\rangle$ as in Eq.(9).
Considering the mixing mechanism, we plot the total decay widths of the mixing states $|1P \frac{3}{2}^{-}\rangle_{1}$ and $|1P \frac{3}{2}^{-}\rangle_{2}$ versus the mixing angle $\theta$ in the range $0^{\circ}\sim30^{\circ}$ in Figs. \ref{Fig 2}-\ref{Fig 3}. When the mixing angle $\theta$ is constrained in $10^{\circ}\sim15^{\circ}$ in Fig. \ref{Fig 3}, the total width of $|1P \frac{3}{2}^{-}\rangle_{2}$ can reach about $30$ MeV, which is consistent with the experimental data. Thus, $\Sigma_{b}(6097)$ can be interpreted as a mixing state of $|\frac{3}{2}^{-},j=1\rangle$ and $|\frac{3}{2}^{-},j=2\rangle$.
It is same for $\Sigma_{c}(2800)$, if the mixing angle equals to $30^{\circ}$ in Fig. \ref{Fig 2}, the total width of $|1P \frac{3}{2}^{-}\rangle_{2}$ can reach $60$ MeV$\sim70$ MeV, which indicates $\Sigma_{c}(2800)$ is possibly a mixing state of $|\frac{3}{2}^{-},j=1\rangle$ and $|\frac{3}{2}^{-},j=2\rangle$.

The theoretical widths of $1P$-wave $\Sigma_{c}$($\frac{5}{2}^{-}$) and $\Sigma_{b}$($\frac{5}{2}^{-}$) which are still missing in experiments, are predicted to be $22.42$ MeV and $18.65$ MeV, respectively(Tables \ref{Table X} and \ref{Table XIII}). It is shown that $\Lambda_{c}\pi^{+}$ and $\Lambda_{b}\pi^{+}$ are the ideal channels to search for these two states. As mentioned in Ref.\cite{GuoL}, the $2S$-wave $\Sigma_{Q}$($\frac{1}{2}^{+}$) and $\Sigma_{Q}$($\frac{3}{2}^{+}$) are also expected to be observed in experiments. The theoretical widths for 2$S$ states $\Sigma_{c1}^{0\rho^{\prime}}$($\frac{1}{2}^{+}$), $\Sigma_{c1}^{0\rho^{\prime}}$($\frac{3}{2}^{+}$), $\Sigma_{b1}^{0\rho^{\prime}}$($\frac{1}{2}^{+}$), and $\Sigma_{b1}^{0\rho^{\prime}}$($\frac{3}{2}^{+}$) are $11\sim13$ MeV(Tables \ref{Table VIII} and \ref{Table XI}), which are relatively narrow. This
implies that these states may easily be observed in future experiments. The decay modes $\Sigma_{c}^{++,+}\pi^{0,+}$ and $\Sigma_{c}^{*++,+}\pi^{0,+}$ provide dominating contributions to the total widths of $\Sigma_{c1}^{0\rho^{\prime}}$($\frac{1}{2}^{+}$) and $\Sigma_{c1}^{0\rho^{\prime}}$($\frac{3}{2}^{+}$), while $\Sigma_{b}^{+,0}\pi^{0,+}$, and $\Sigma_{b}^{*+,0}\pi^{0,+}$ are the dominating decay channels for $\Sigma_{b1}^{0\rho^{\prime}}$($\frac{1}{2}^{+}$) and $\Sigma_{b1}^{0\rho^{\prime}}$($\frac{3}{2}^{+}$) states.

\begin{large}
\textbf{3.3 $\Omega_{Q}$ states}
\end{large}
\begin{table*}[h]
\begin{ruledtabular}\caption{Experimental information of $\Omega_{Q}$(Q=c,b) baryons, and the predictions by quark model and $^{3}P_{0}$ model. All values are in units of MeV}
\label{Table III}
\begin{tabular}{c c c c c |c c |c}
\multicolumn{5}{c}{Experimental information\cite{article2B,article2A}} & \multicolumn{2}{c}{Quark model\cite{GuoL}} &$^{3}P_{0}$ model \\
States  & $J^{P}$ &Mass  &Width &Decay channels  &  $J^{P}$    &  Mass  & width\\ \hline
$\Omega_{c}^{0}$&$\frac{1}{2}^{+}$&2695.2$\pm$1.7&-&- &$\frac{1}{2}^{+}$(1S)& 2699&-\\
$\Omega_{c}(2770)^{0}$&$\frac{3}{2}^{+}$&2765.9$\pm$2.0&-&- &$\frac{3}{2}^{+}$(1S)& 2762& -\\
$\Omega_{c}(3000)^{0}$&$?^{?}$&3000.4$\pm$0.2$\pm$0.1$\pm$0.3&4.5$\pm$0.6$\pm$0.3&$\Xi^{+}_{c}K^{-}$ &1P($\frac{1}{2}^{-}$)$_{j=1}$&3045& $4.00\sim$5.00 \\
$\Omega_{c}(3050)^{0}$&$?^{?}$&3050.2$\pm$0.1$\pm$0.1$\pm$0.3&$<1.2$&$\Xi^{+}_{c}K^{-}$  &1P($\frac{3}{2}^{-}$)$_{j=1}$&3062& $<$1.00\\
$\Omega_{c}(3065)^{0}$&$?^{?}$&3065.6$\pm$0.1$\pm$0.3$\pm$0.3&3.5$\pm$0.4$\pm$0.2&$\Xi^{+}_{c}K^{-}$ &1P($\frac{3}{2}^{-}$)$_{j=2}$&3039& 1.00\\
$\Omega_{c}(3090)^{0}$&$?^{?}$&3090.2$\pm$0.3$\pm$0.5$\pm$0.3&8.7$\pm$1.0$\pm$0.8&$\Xi^{+}_{c}K^{-}$ &1P($\frac{5}{2}^{-}$)$_{j=2}$&3067& 2.50\\
$\Omega_{c}(3120)^{0}$&$?^{?}$&3119.1$\pm$0.3$\pm$0.9$\pm$0.3&$<2.6$&$\Xi^{+}_{c}K^{-}$ &$\frac{1}{2}^{+}$(2S)&3150& 5.95\\
$\Omega_{c}(3185)^{0}$&$?^{?}$&3185.1$\pm$1.7$^{+7.4}_{-0.9}\pm$0.2&50$\pm$7$^{+10}_{-20}$&$\Xi^{+}_{c}K^{-}$ &$\frac{3}{2}^{+}$(2S)&3197& 78.16\\
$\Omega_{c}(3327)^{0}$&$?^{?}$&3327.1$\pm$1.2$^{+0.1}_{-1.3}\pm$0.2&20$\pm$5$^{+13}_{-1}$&$\Xi^{+}_{c}K^{-}$ &$\frac{3}{2}^{+}$(1D)&3313& 25.20\\
$\Omega_{b}(6316)^{-}$&$?^{?}$&6315.64$\pm$0.31$\pm$0.07$\pm$0.50&$<2.8(4.2)$&$\Xi_{b}^{0}K^{-}$ &1P($\frac{1}{2}^{-}$)$_{j=1}$&6329& 1.00$\sim$2.00\\
$\Omega_{b}(6330)^{-}$&$?^{?}$&6330.30$\pm$0.28$\pm$0.07$\pm$0.50&$<3.1(4.7)$&$\Xi_{b}^{0}K^{-}$ &1P($\frac{3}{2}^{-}$)$_{j=1}$&6336& 0.23\\
$\Omega_{b}(6340)^{-}$&$?^{?}$&6339.71$\pm$0.26$\pm$0.05$\pm$0.50&$<1.5(1.8)$&$\Xi_{b}^{0}K^{-}$ &1P($\frac{3}{2}^{-}$)$_{j=2}$&6326& 0.05\\
$\Omega_{b}(6350)^{-}$&$?^{?}$&6349.88$\pm$0.35$\pm$0.05$\pm$0.50&$<2.8(3.2)$&$\Xi_{b}^{0}K^{-}$  &1P($\frac{5}{2}^{-}$)$_{j=2}$&6339& 0.55\\
\end{tabular}
\end{ruledtabular}
\end{table*}

 As for the newly observed $\Omega_{c}(3185)$ and $\Omega_{c}(3327)$, their masses are consistent with the predictions for 2$S$($\frac{3}{2}^{+}$) and 1$D$ states by quark model\cite{GuoL}. As a 2$S$-wave $\Omega_{c1}^{0\lambda^{\prime}}$($\frac{3}{2}^{+}$) state, the theoretical width of $\Omega_{c}(3185)$ is 78.16 MeV(Tables \ref{Table III} and \ref{Table XIV}). This value is close to the experimental data 50$\pm$7$^{+10}_{-20}$ MeV. Thus, it is reasonable to assign $\Omega_{c}(3185)$ as a 2$S$($\frac{3}{2}^{+}$) state. The $\Omega_{c}(3327)$ is discovered in the decay channel $\Xi_{c}^{+}K^{-}$ with a total width being 20$\pm$5$^{+13}_{-1}$ Mev\cite{3185}. From Table \ref{Table XVII}, we can see the decay channels and total widths of 1D-wave $\frac{1}{2}^{+}$ and $\frac{3}{2}^{+}$ states are both compatible with experimental data. However, the predicted mass for the latter, 3313 MeV\cite{GuoL}, is closer to experiments. Therefore, $\Omega_{c}(3327)$ is possibly a 1D($\frac{3}{2}^{+}$) state.

 According to the mass spectrum\cite{GuoL}, the previously observed baryons $\Omega_{c}(3120)$, ($\Omega_{c}(3000)$, $\Omega_{c}(3050)$) and ($\Omega_{c}(3065)$, $\Omega_{c}(3090)$) were suggested as a $2S$($\frac{1}{2}^{+}$) state, the $1P$ doublets ($\frac{1}{2}^{-}$,$\frac{3}{2}^{-}$)$_{j=1}$ and ($\frac{3}{2}^{-}$,$\frac{5}{2}^{-}$)$_{j=2}$, respectively. As a $2S$-wave $\Omega_{c1}^{0\rho^{\prime}}$($\frac{1}{2}^{+}$) state, the theoretical width of $\Omega_{c}(3120)$ is $5.95$ MeV(Table \ref{Table XIV}), which is slightly larger than the experimental data(Table \ref{Table III}). Given the theoretical uncertainty, it is reasonable to interpret $\Omega_{c}(3120)$ as a $2S$($\frac{1}{2}^{+}$) state. This also means that $\Omega_{c}(3120)$ and $\Omega_{c}(3185)$ are a 2$S$ doublet ($\frac{1}{2}^{+}$, $\frac{3}{2}^{+}$). It is shown in Tables \ref{Table XIV}-\ref{Table XVI} that the predicted widths for the five $1P$-wave states, $\Omega_{c0}^{1}$($\frac{1}{2}^{-}$), $\Omega_{c1}^{1}$($\frac{1}{2}^{-}$), $\Omega_{c1}^{1}$($\frac{3}{2}^{-}$), $\Omega_{c2}^{1}$($\frac{3}{2}^{-}$) and $\Omega_{c2}^{1}$($\frac{5}{2}^{-}$) are $364$ MeV, $2.18\times10^{-11}$ MeV, $2.85\times10^{-10}$ MeV, $1.08$ MeV and $2.50$ MeV, respectively. In comparison with the experimental data, the assignments for $\Omega_{c}(3065)$ and $\Omega_{c}(3090)$ as a doublet ($\frac{3}{2}^{-}$,$\frac{5}{2}^{-}$)$_{j=2}$ is reasonable. However, the predicted widths of the doublet ($\frac{1}{2}^{-}$,$\frac{3}{2}^{-}$)$_{j=1}$ are very tiny, which means the above assignments for $\Omega_{c}(3000)$ and $\Omega_{c}(3050)$ is unreasonable. In Ref.\cite{3P0161}, these two $\Omega_{c}$ baryons were suggested to be the $1D$-wave states. However, the predicted masses in quark model for $1D$-wave $\Omega_{c}$ are much larger than the experimental data, which indicates $1D$-wave states are not good candidates for $\Omega_{c}(3000)$ and $\Omega_{c}(3050)$. Again, if we consider the mixing mechanism, $\Omega_{c}(3000)$ and $\Omega_{c}(3050)$ still can be described as the $1P$-wave states.
\begin{figure}[h]
\begin{minipage}[h]{0.45\linewidth}
\centering
\includegraphics[height=5cm,width=7cm]{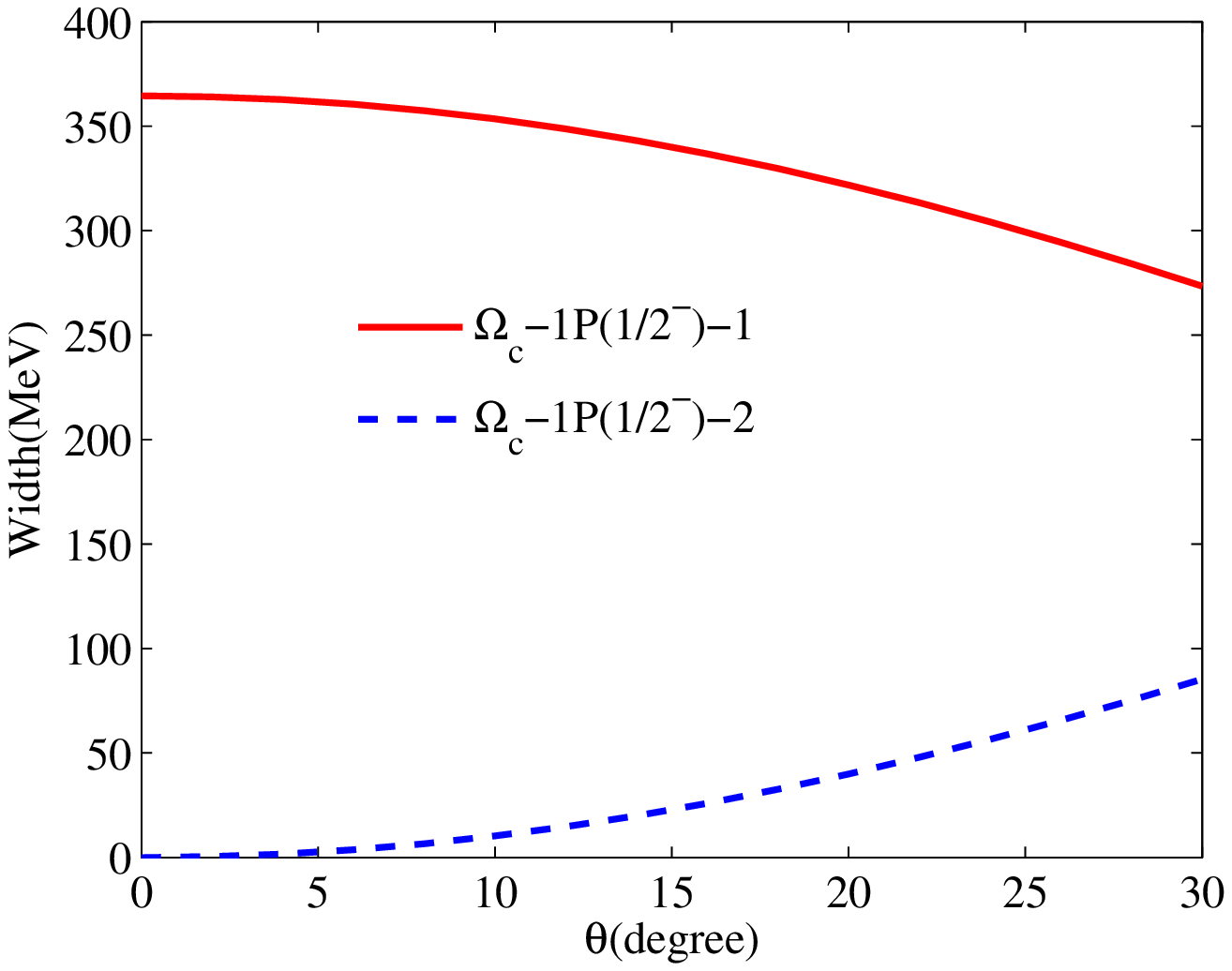}
\caption{The total decay widths of $|1P\frac{1}{2}^{-}\rangle_{1}$ and $|1P\frac{1}{2}^{-}\rangle_{2}$ for $\Omega_{c}$ as functions
of the mixing angle $\theta$ in the range $0^{\circ}\sim30^{\circ}$.}
\label{Fig 4}
\end{minipage}
\hfill
\begin{minipage}[h]{0.45\linewidth}
\centering
\includegraphics[height=5cm,width=7cm]{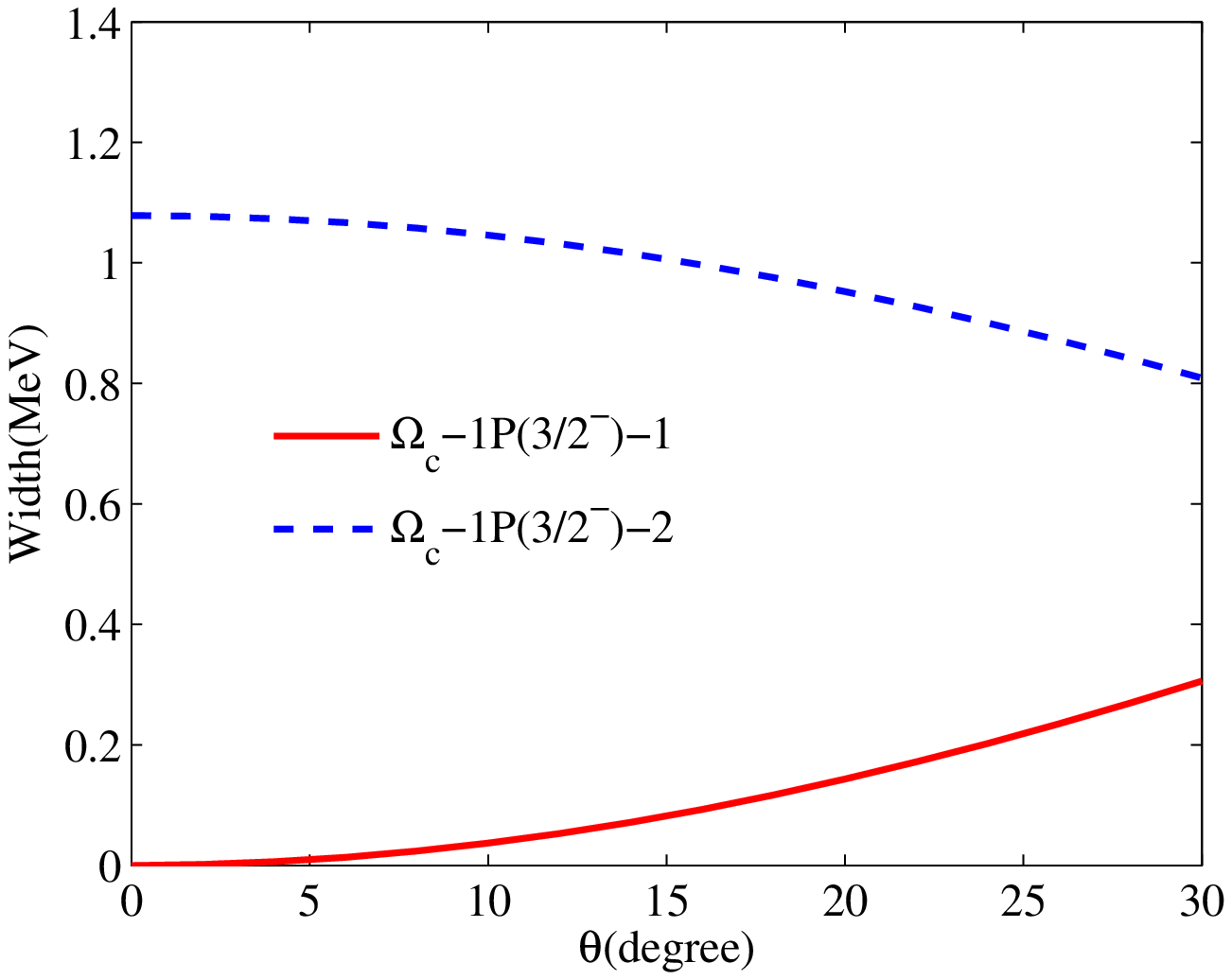}
\caption{The total decay widths of $|1P\frac{3}{2}^{-}\rangle_{1}$ and $|1P\frac{3}{2}^{-}\rangle_{2}$ for $\Omega_{c}$ as functions
of the mixing angle $\theta$ in the range $0^{\circ}\sim30^{\circ}$.}
\label{Fig 5}
\end{minipage}
\end{figure}
\begin{figure}[h]
\begin{minipage}[h]{0.45\linewidth}
\centering
\includegraphics[height=5cm,width=7cm]{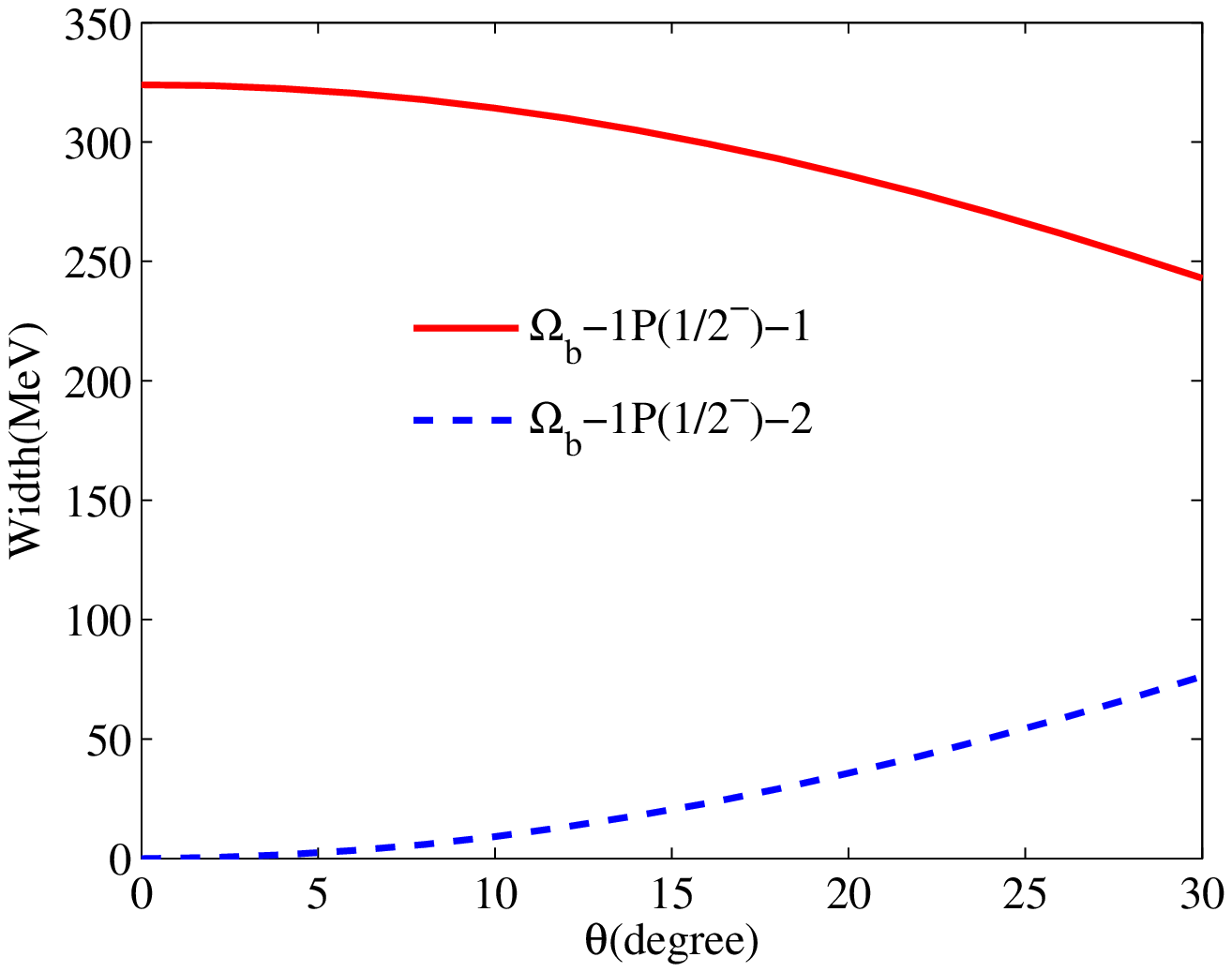}
\caption{The total decay widths of $|1P\frac{1}{2}^{-}\rangle_{1}$ and $|1P\frac{1}{2}^{-}\rangle_{2}$ for $\Omega_{b}$ as functions
of the mixing angle $\theta$ in the range $0^{\circ}\sim30^{\circ}$.}
\label{Fig 6}
\end{minipage}
\hfill
\begin{minipage}[h]{0.45\linewidth}
\centering
\includegraphics[height=5cm,width=7cm]{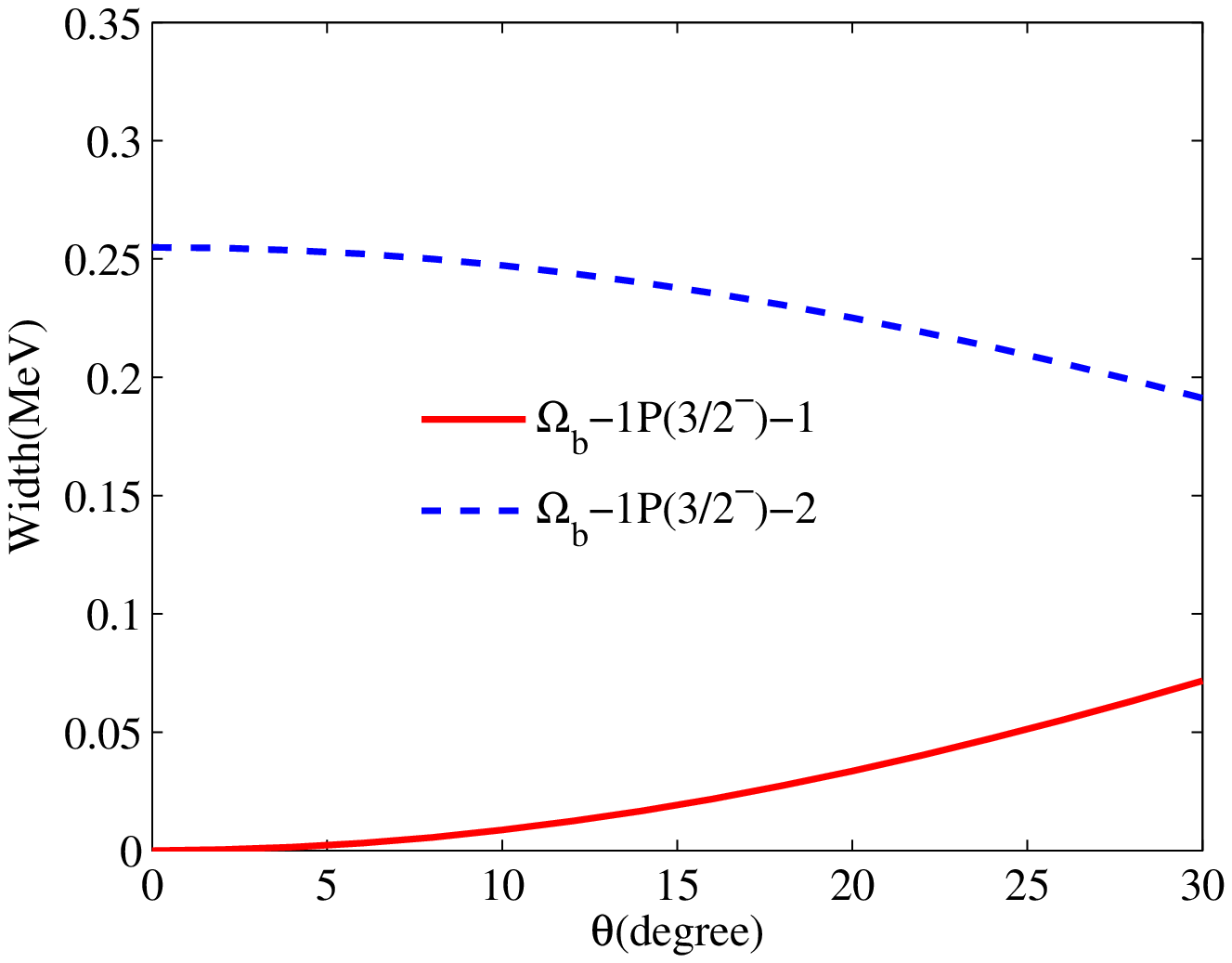}
\caption{The total decay widths of $|1P\frac{3}{2}^{-}\rangle_{1}$ and $|1P\frac{3}{2}^{-}\rangle_{2}$ for $\Omega_{b}$ as functions
of the mixing angle $\theta$ in the range $0^{\circ}\sim30^{\circ}$.}
\label{Fig 7}
\end{minipage}
\end{figure}

We plot the total widths of the mixing states versus the mixing angle $\theta$ in the range $0^{\circ}\sim30^{\circ}$ in Figs. \ref{Fig 4}-\ref{Fig 5}. It is shown that three mixing states $|1P \frac{1}{2}^{-}\rangle_{2}$, $|1P \frac{3}{2}^{-}\rangle_{1}$, and $|1P \frac{3}{2}^{-}\rangle_{2}$ belong to the narrow resonances. If the mixing angle $\theta$ is constrained in a small value in Fig. \ref{Fig 4}, the total width of $|1P \frac{1}{2}^{-}\rangle_{2}$ reaches 4$\sim$5 MeV. In addition, the values for $|1P \frac{3}{2}^{-}\rangle_{1}$ and $|1P \frac{3}{2}^{-}\rangle_{2}$ in Fig. \ref{Fig 5} are all lower than $1$ MeV. These results are roughly compatible with the experimental data. Thus, we tentatively assign the $\Omega_{c}(3000)$, $\Omega_{c}(3050)$, and $\Omega_{c}(3065)$ as the mixing states $|1P \frac{1}{2}^{-}\rangle_{2}$, $|1P \frac{3}{2}^{-}\rangle_{1}$ and $|1P \frac{3}{2}^{-}\rangle_{2}$, respectively and assign $\Omega_{c}(3090)$ as a pure $1P$($\frac{5}{2}^{-}$) state.

As for the narrow resonances, $\Omega_{b}(6316)$, $\Omega_{b}(6330)$, $\Omega_{b}(6340)$, and $\Omega_{b}(6350)$, their situation is very similar with the $1P$-wave $\Omega_{c}$ states. After considering state mixing, the total width are plotted in Figs. \ref{Fig 6}-\ref{Fig 7}.  From these figures, we can obtain the similar conclusions with $\Omega_{c}$ baryons, that $\Omega_{b}(6316)$, $\Omega_{b}(6330)$, $\Omega_{b}(6340)$, and $\Omega_{b}(6350)$ can be respectively described as three mixing states $|1P \frac{1}{2}^{-}\rangle_{2}$, $|1P \frac{3}{2}^{-}\rangle_{1}$, $|1P \frac{3}{2}^{-}\rangle_{2}$, and a pure state $1P$($\frac{5}{2}^{-}$). In Refs. \cite{WL,3P0Z3}, they also obtained the same conclusions with ours.

Up to now, the $2S$ $\Omega_{c}$ doublet ($\frac{1}{2}^{+}$, $\frac{3}{2}^{+}$) have been observed. However, their $\Omega_{b}$ partners $\Omega_{b}$($\frac{1}{2}^{+}$) and $\Omega_{b}$($\frac{3}{2}^{+}$) are still missing in experiments. The results in Table \ref{Table XVIII} show that the radial excited mode of these two states may be the $\rho^{\prime}$ excitations with ($n_{\rho}$, $n_{\lambda}$)=($1$, $0$). Theoretical widths for these two states are $5.06$ MeV and $4.63$ MeV, respectively. They have the similar decay behaviors, both of them dominantly decay into $\Xi_{b}^{0}K^{-}$ and $\Xi_{b}^{-}K^{0}$.
\begin{figure}[h]
\begin{minipage}[h]{0.45\linewidth}
\centering
\includegraphics[height=5cm,width=7cm]{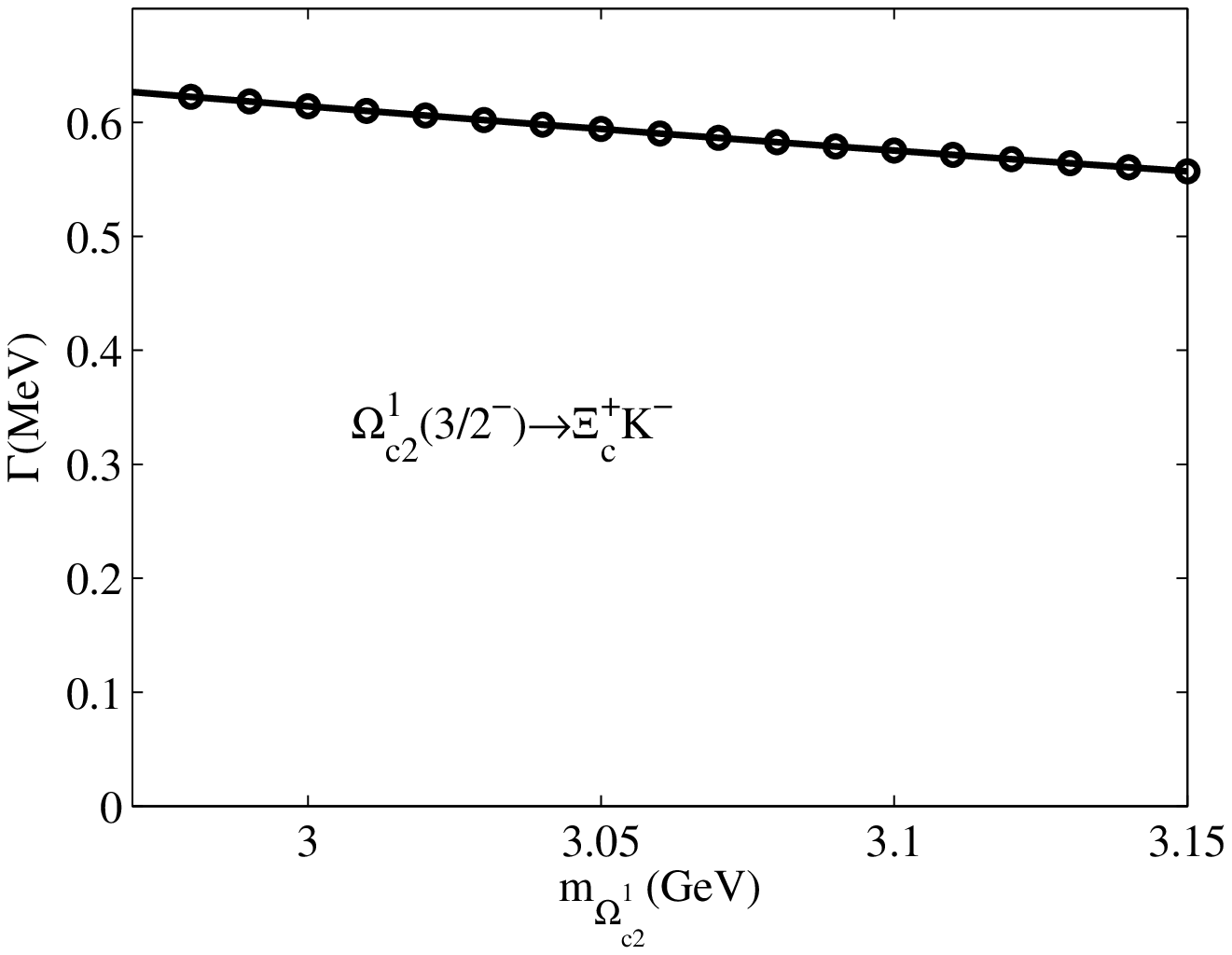}
\caption{The variation of the decay width of $\Omega_{c2}^{1}$($\frac{3}{2}^{-}$)$\rightarrow$$\Xi_{c}^{+}K^{-}$ with the mass of $\Omega_{c2}^{1}$($\frac{3}{2}^{-}$).}
\label{Fig 8}
\end{minipage}
\hfill
\begin{minipage}[h]{0.45\linewidth}
\centering
\includegraphics[height=5cm,width=7cm]{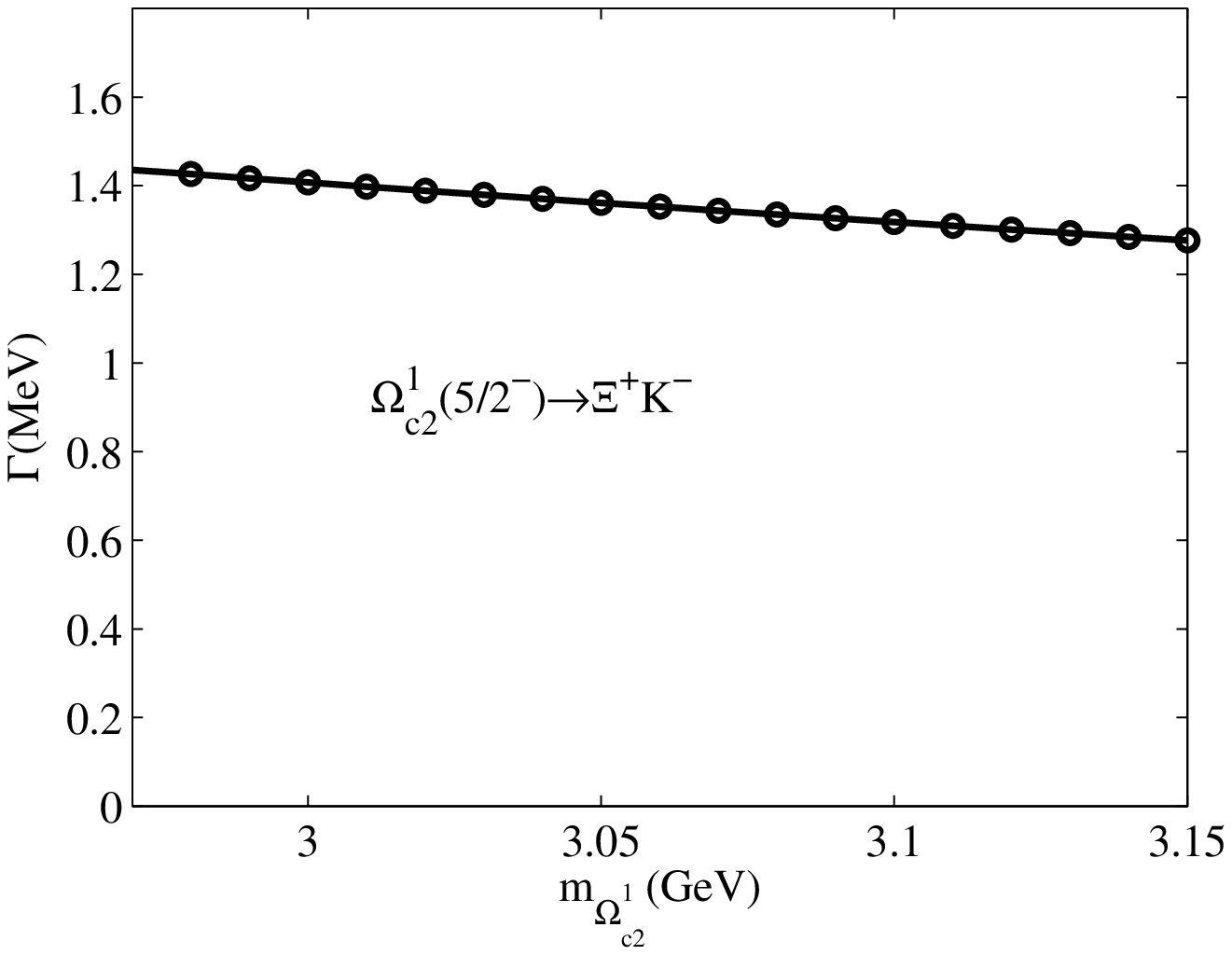}
\caption{The variation of the decay width of $\Omega_{c2}^{1}$($\frac{5}{2}^{-}$)$\rightarrow$$\Xi_{c}^{+}K^{-}$ with the mass of $\Omega_{c2}^{1}$($\frac{5}{2}^{-}$).}
\label{Fig 9}
\end{minipage}
\end{figure}
\begin{figure}[h]
\begin{minipage}[h]{0.45\linewidth}
\centering
\includegraphics[height=5cm,width=7cm]{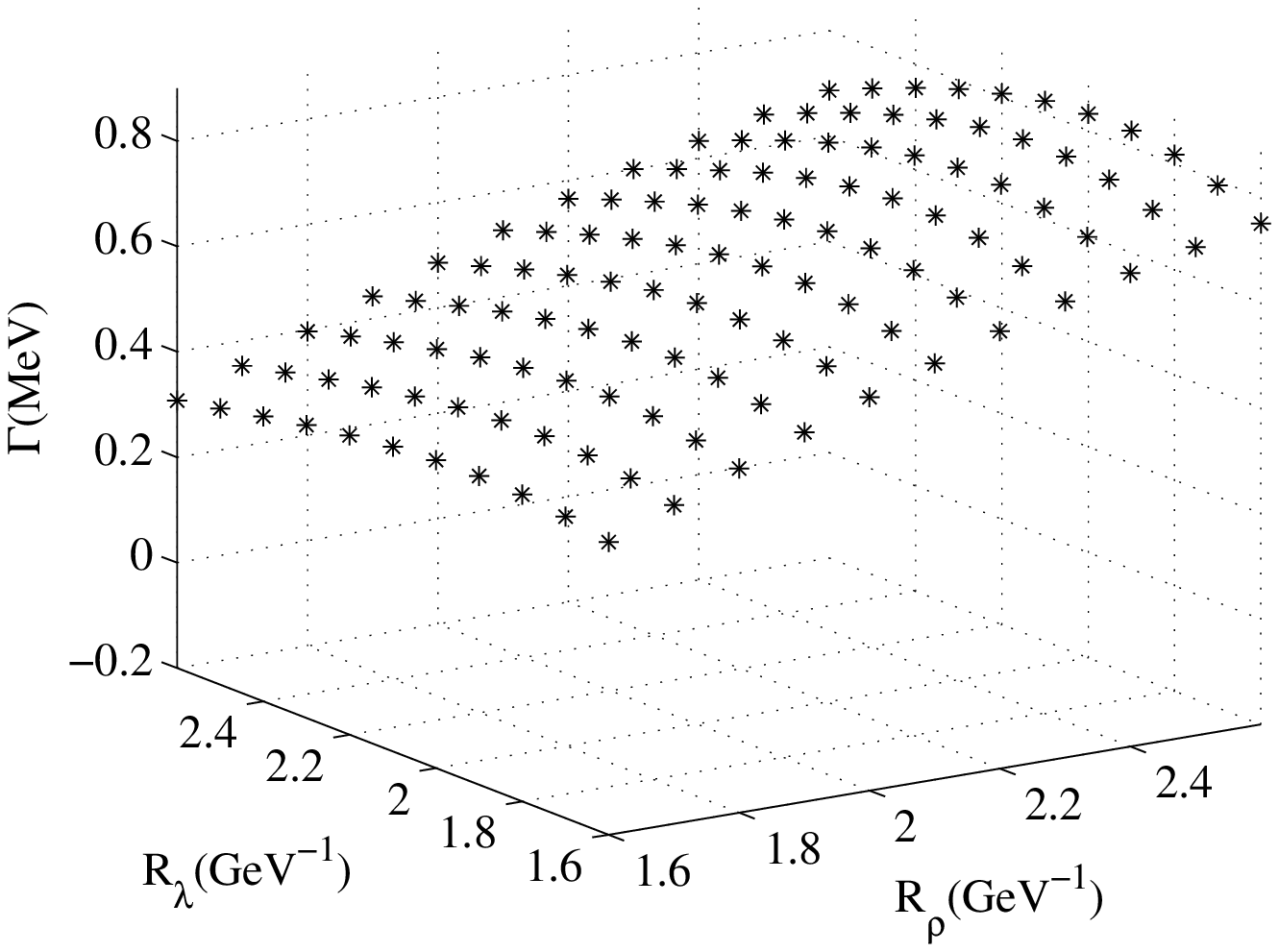}
\caption{The variation of the decay width of $\Omega_{c2}^{1}$($\frac{3}{2}^{-}$)$\rightarrow$$\Xi_{c}^{+}K^{-}$ with $R_{\rho}$ and $R_{\lambda}$.}
\label{Fig 10}
\end{minipage}
\hfill
\begin{minipage}[h]{0.45\linewidth}
\centering
\includegraphics[height=5cm,width=7cm]{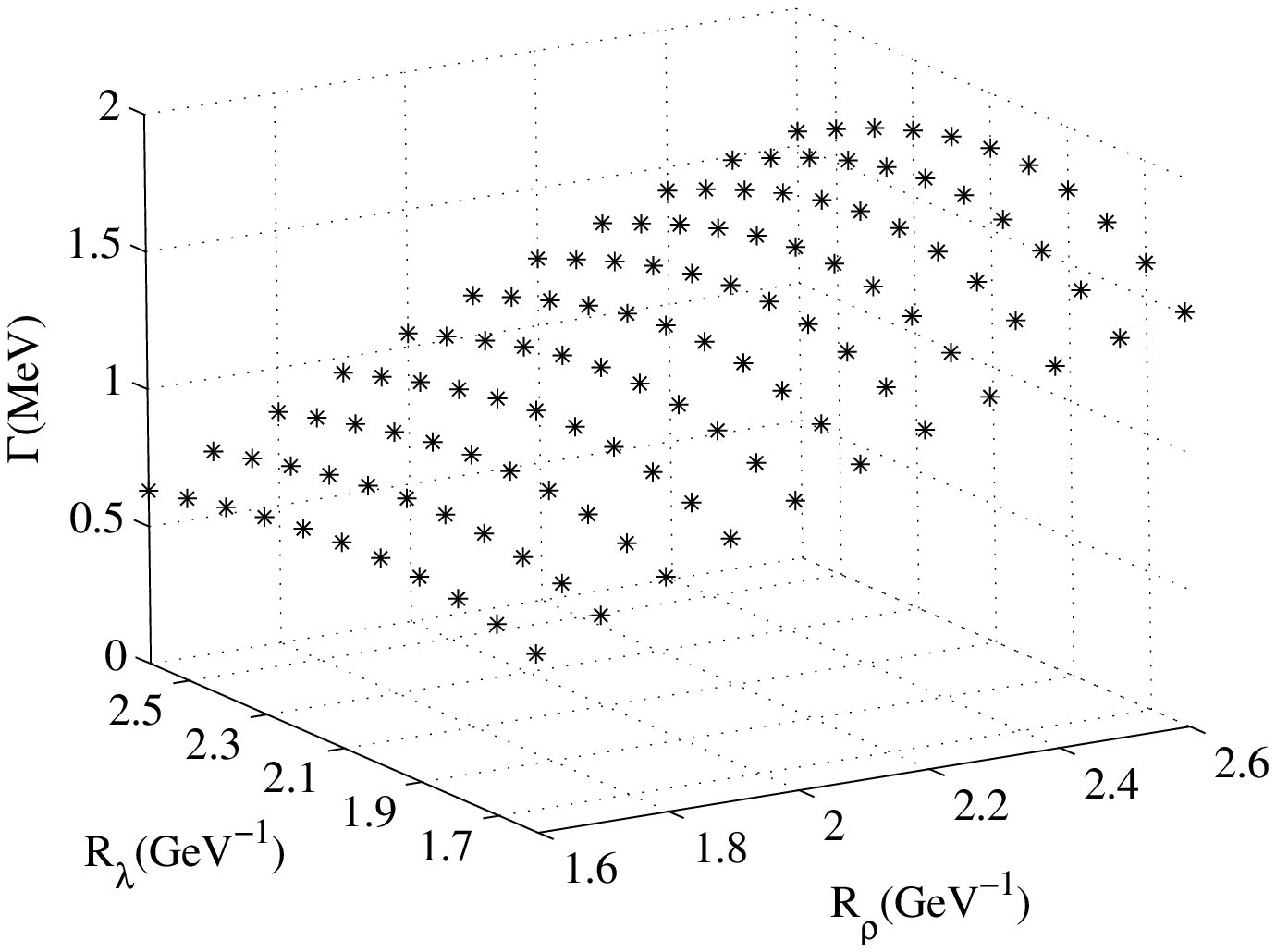}
\caption{The variation of the decay width of $\Omega_{c2}^{1}$($\frac{5}{2}^{-}$)$\rightarrow$$\Xi_{c}^{+}K^{-}$ with $R_{\rho}$ and $R_{\lambda}$.}
\label{Fig 11}
\end{minipage}
\end{figure}

As for the uncertainties of $^{3}P_{0}$ model, it arise from the quark pair creation strength $\gamma$, the masses of hadrons, and the SHO wave function parameter $R$. The parameter of $\gamma$ describes the strength of quark-antiquark pair creation from the vacuum, and it is usually taken as the universal value of 13.4 which is fitted according to experimental data\cite{3P0M01A,3P05,3P010}. As for the masses of hadrons and scale parameters $R$, we illustrate the dependence of the results on them in Figs. \ref{Fig 8}-\ref{Fig 11}, using two typical decay channels $\Omega_{c2}^{1}$($\frac{3}{2}^{-}$)$\rightarrow$$\Xi_{c}^{+}K^{-}$ and $\Omega_{c2}^{1}$($\frac{5}{2}^{-}$)$\rightarrow$$\Xi_{c}^{+}K^{-}$. It is shown in Figs. \ref{Fig 8}-\ref{Fig 9} that the masses of initial heavy baryons have limited influence on the results. However, the predicted width changes $2\sim3$ times when the parameters $R_{\rho}$ and $R_{\lambda}$ of $\Omega_{c2}^{1}$ states change from $1.6$ GeV$^{-1}$ to $2.6$ GeV$^{-1}$(Figs. \ref{Fig 10}-\ref{Fig 11}). This indicates that the uncertainties mainly originate from parameters $R$. As discussed at the beginning of Section 3, we carefully take their values which are either fixed by fitting experimental data\cite{3P0M01A,3P0M01B,3P0M01C} or determined by solving the Schr\"{o}dinger equation\cite{3P05,3P010,Bpara,3P017,3P018}. It is difficult to calculate the exact uncertainties of results originated from $R$ because the uncertainties of $R$ is unknown. It was stated in Ref. \cite{3P05} that the maximum deviation of theoretical results may be a factor of $2\sim3$ off the experimental width. Even with the above uncertainty, the $^{3}P_{0}$ model is still the most systematic, effective, and widely used framework to study the baryon strong decays. More detailed analysis about the uncertainties of the results in the $^{3}P_{0}$ decay model can be found in Refs. \cite{3P0M01A,3P05}.

\begin{Large}
\textbf{4 Conclusions}
\end{Large}

In this work, we have systematically investigated strong decay behaviors of the single heavy baryons $\Lambda_{Q}$, $\Sigma_{Q}$, and $\Omega_{Q}$.
A number of experimental states without spin-parity assignments are successfully distinguished. For example, the $\Lambda_{c}(2940)$, $\Lambda_{c}(2860)$, $\Lambda_{c}(2880)$, $\Lambda_{b}(6070)$ are suggested to be the $2P$($\frac{1}{2}^{-}$), $1D$($\frac{1}{2}^{+}$), $1D$($\frac{3}{2}^{+}$) and $2S$($\frac{1}{2}^{+}$) states, respectively. The $\Sigma_{c}(2800)$ and $\Sigma_{b}(6097)$ may be the pure quark model state $1P$($\frac{3}{2}^{-}$)$_{j=2}$. Another possible interpretation is that each of them is the mixing state of $|\frac{3}{2}^{-},j=1\rangle$ and $|\frac{3}{2}^{-},j=2\rangle$. The $\Omega_{c}(3120)$ and $\Omega_{c}(3185)$ are supported as the 2S doublet ($\frac{1}{2}^{+}$, $\frac{3}{2}^{+}$) and $\Omega_{c}(3090)$ is assigned as the pure quark model state $1P$($\frac{5}{2}^{-}$). Model predictions support assigning the recently observed $\Omega_{c}(3327)$ as a 1D($\frac{3}{2}^{+}$) state. As for $\Omega_{c}(3000)$, $\Omega_{c}(3050)$ and $\Omega_{c}(3065)$, they can be interpreted as the mixing states $|1P \frac{1}{2}^{-}\rangle_{2}$, $|1P \frac{3}{2}^{-}\rangle_{1}$, and $|1P \frac{3}{2}^{-}\rangle_{2}$, respectively. The $\Omega_{b}(6316)$, $\Omega_{b}(6330)$, $\Omega_{b}(6340)$, and $\Omega_{b}(6350)$ can be described as the mixing states $|1P \frac{1}{2}^{-}\rangle_{2}$, $|1P \frac{3}{2}^{-}\rangle_{1}$, $|1P \frac{3}{2}^{-}\rangle_{2}$, and a pure quark model state $1P$($\frac{5}{2}^{-}$), respectively.

A number of single heavy barons which have good potentials to be discovered in forthcoming experiments are predicted and some valuable clues for searching for these missing baryons are suggested by $^{3}P_{0}$ decay model. For $\Lambda_{Q}$ systems, we suggest to search for the $2P$-wave state $\Lambda_{c1}^{1\lambda^{\prime} }$($\frac{3}{2}^{-}$) in the decay channels $\Sigma_{c}^{*+,0}\pi^{0,+}$, $D^{*0}p$ and $D^{*+}n$. For $\Lambda_{b}$ spin-doublet ($\frac{1}{2}^{-}$, $\frac{3}{2}^{-}$)$_{j=1}$, they are most likely to be found in decay channels $\Sigma_{b}^{+,0,-}\pi^{-,0,+}$ and $\Sigma_{b}^{*+,0,-}\pi^{-,0,+}$, respectively.
 As mentioned in Section 3.1, $\Lambda_{c}\pi^{+}$ and $\Lambda_{b}\pi^{+}$ are the ideal decay channels to find the $1P$-wave $\Sigma_{c}$($\frac{5}{2}^{-}$) and $\Sigma_{b}$($\frac{5}{2}^{-}$) states. The $2S$-wave $\Sigma_{c1}^{0\rho^{\prime}}$($\frac{1}{2}^{+}$) and $\Sigma_{c1}^{0\rho^{\prime}}$($\frac{3}{2}^{+}$) have good potentials to be observed in the $\Sigma_{c}^{++,+}\pi^{0,+}$ and $\Sigma_{c}^{*++,+}\pi^{0,+}$ decay channels, while $\Sigma_{b}^{+,0}\pi^{0,+}$, and $\Sigma_{b}^{*+,0}\pi^{0,+}$ are the dominating decay modes for $\Sigma_{b1}^{0\rho^{\prime}}$($\frac{1}{2}^{+}$) and $\Sigma_{b1}^{0\rho^{\prime}}$($\frac{3}{2}^{+}$) states.
Finally, we suggest to hunt for the $2S$-wave doublet $\Omega_{b}$($\frac{1}{2}^{+}$, $\frac{3}{2}^{+}$) in the $\Xi_{b}^{0}K^{-}$ and $\Xi_{b}^{-}K^{0}$ decay channels.

\begin{Large}
Acknowledgments
\end{Large}

We would like to thank Si-Qiang Luo and Xiang Liu for their valuable discussions. This project is supported by the Natural Science Foundation of HeBei Province, Grant Number A2018502124.

\begin{center}
\begin{Large}
\textbf{Appendix A: Two-body strong decays of $\Lambda_{Q}$, $\Sigma_{Q}$ and $\Omega_{Q}$}
\end{Large}
\end{center}
\begin{table*}[h]
\begin{ruledtabular}\caption{Two-body strong decays of $\Lambda_{c}$ in MeV}
\label{Table IV}
\begin{tabular}{c c c c c c c c }
N&1&2&3&4&5&6&7 \\  \hline
Notations & $\Lambda_{c0}^{0\lambda^{\prime}}$ & $\Lambda_{c0}^{0\rho^{\prime} }$ &$\Lambda_{c1}^{1}$ & $\Lambda_{c1}^{1\lambda^{\prime} }$ & $\Lambda_{c1}^{1\rho^{\prime} }$ & $\Lambda_{c1}^{1}$ & $\Lambda_{c1}^{1\lambda^{\prime} }$ \\ \hline
Assignments & $\frac{1}{2}^{+}$ (2S)& $\frac{1}{2}^{+}$ (2S)& $\frac{1}{2}^{-}$ (1P)& $\frac{1}{2}^{-}$ (2P)& $\frac{1}{2}^{-}$ (2P)&$\frac{3}{2}^{-}$ (1P) & $\frac{3}{2}^{-}$ (2P) \\ \hline
Mass& 2764 & 2764 &2592 & 2988 & 2988 & 2628 & 3013 \\ \hline
$\Sigma_{c}^{++}$$\pi^{-}$&5.04&1.49&2.72&2.34&6.84&3.09$\times10^{-3}$&8.50$\times10^{-2}$\\    \hline
$\Sigma_{c}^{+}$$\pi^{0}$&5.14&1.50&6.00&2.34&6.77&3.93$\times10^{-3}$&8.70$\times10^{-2}$ \\    \hline
$\Sigma_{c}^{0}$$\pi^{+}$&5.04&1.49&2.72&2.34&6.84&3.09$\times10^{-3}$&8.50$\times10^{-2}$ \\    \hline
$\Sigma_{c}^{*++}$$\pi^{-}$&3.33&1.26&-&5.20$\times10^{-2}$&5.74&-&2.17 \\    \hline
$\Sigma_{c}^{*+}$$\pi^{0}$&3.46&1.30&-&5.70$\times10^{-2}$&5.80&-&2.17 \\    \hline
$\Sigma_{c}^{*0}$$\pi^{+}$&3.33&1.26&-&0.052&5.74&-&2.17 \\    \hline
$D^{*0}$$p$&-&-&-&2.67&2.81&-&3.31 \\    \hline
$D^{*+}$$n$&-&-&-&2.55&2.80&-&3.23\\    \hline
$\Gamma_{total}$&25.34&8.30&11.44&12.40&43.34&1.00$\times10^{-3}$&13.31 \\
\end{tabular}
\end{ruledtabular}
\end{table*}
\begin{table*}[h]
\begin{ruledtabular}\caption{Two-body strong decays of $\Lambda_{c}$ in MeV}
\label{Table V}
\begin{tabular}{c c c c c c c c  }
N&1&2&3&4&5&6&7  \\  \hline
Notations & $\Lambda_{c1}^{1\rho^{\prime}}$ & $\Lambda_{c2}^{2}$ &$\Lambda_{c2}^{2\lambda^{\prime} }$ & $\Lambda_{c2}^{2\rho^{\prime} }$ & $\Lambda_{c2}^{2}$ & $\Lambda_{c2}^{2\lambda^{\prime} }$ & $\Lambda_{c2}^{2\rho^{\prime} }$ \\ \hline
 Assignments& $\frac{3}{2}^{-}$ (2P)& $\frac{3}{2}^{+}$ (1D)& $\frac{3}{2}^{+}$ (2D)& $\frac{3}{2}^{+}$ (2D)& $\frac{5}{2}^{+}$ (1D) & $\frac{5}{2}^{+}$ (2D)& $\frac{5}{2}^{+}$ (2D) \\ \hline
Mass &3013 & 2856 & 3220 & 3220 &2882 & 3234 & 3234 \\ \hline
$\Sigma_{c}^{++}$$\pi^{-}$&6.47&2.69&3.10&8.90$\times10^{-2}$&5.60$\times10^{-2}$&8.40$\times10^{-1}$& 7.47$\times10^{-1}$ \\    \hline
$\Sigma_{c}^{+}$$\pi^{0}$&6.51&2.71&3.11&8.70$\times10^{-2}$&5.70$\times10^{-2}$&8.43$\times10^{-1}$&7.50$\times10^{-1}$ \\    \hline
$\Sigma_{c}^{0}$$\pi^{+}$&6.47&2.69&3.10&8.90$\times10^{-2}$&5.60$\times10^{-2}$&8.39$\times10^{-1}$&7.47$\times10^{-1}$ \\    \hline
$\Sigma_{c}^{*++}$$\pi^{-}$&12.4&0.313&1.36&8.98$\times10^{-1}$&2.34&3.93&6.53$\times10^{-1}$ \\    \hline
$\Sigma_{c}^{*+}$$\pi^{0}$&12.4&0.318&1.37&9.01$\times10^{-2}$&2.37&3.93&6.52$\times10^{-1}$ \\    \hline
$\Sigma_{c}^{*0}$$\pi^{+}$&12.4&0.313&1.36&8.98$\times10^{-2}$&2.34&3.93&6.53$\times10^{-1}$ \\    \hline
$\Lambda_{c}$$\omega$&-&-&2.53&2.57&-&2.81&2.70 \\    \hline
$\Xi_{c}^{'+}$$K^{0}$&-&-&1.58$\times10^{-1}$&7.50$\times10^{-2}$&-&2.67$\times10^{-3}$&4.84$\times10^{-3}$ \\    \hline
$\Xi_{c}^{'0}$$K^{+}$&-&-&1.58$\times10^{-1}$&7.50$\times10^{-2}$&-&2.67$\times10^{-3}$&4.84$\times10^{-3}$ \\    \hline
$D^{*0}$$p$&2.69&-&2.90&1.37$\times10^{-1}$&-&3.04&1.66$\times10^{-1}$ \\    \hline
$D^{*+}$$n$&2.73&-&2.87&1.30$\times10^{-1}$&-&3.00&1.58$\times10^{-1}$\\    \hline
$\Xi_{c}^{'*+}$$K^{0}$&-&-&1.30$\times10^{-2}$&8.84$\times10^{-3}$&-&9.90$\times10^{-2}$&6.00$\times10^{-2}$ \\    \hline
$\Xi_{c}^{'*0}$$K^{+}$&-&-&1.30$\times10^{-2}$&8.84$\times10^{-3}$&-&9.90$\times10^{-2}$&6.00$\times10^{-2}$ \\    \hline
$\Gamma_{total}$&62.07&9.03&22.04&5.97&7.22&23.37&7.36 \\
\end{tabular}
\end{ruledtabular}
\end{table*}
\begin{table*}[h]
\begin{ruledtabular}\caption{Two-body strong decays of $\Lambda_{b}$ in MeV}
\label{Table VI}
\begin{tabular}{c c c c c c c c }
N&1&2&3&4&5&6 \\  \hline
Notations & $\Lambda_{b0}^{1\lambda^{\prime} }$&$\Lambda_{b0}^{1\rho^{\prime} }$&$\Lambda_{b1}^{1\lambda^{\prime} }$&$\Lambda_{b1}^{1\rho^{\prime} }$&$\Lambda_{b1}^{1\lambda^{\prime} }$&$\Lambda_{b1}^{1\rho^{\prime} }$ \\ \hline
Assignments& $\frac{1}{2}^{+}$(2S) & $\frac{1}{2}^{+}$ (2S)& $\frac{1}{2}^{-}$ (2P)& $\frac{1}{2}^{-}$ (2P)& $\frac{3}{2}^{-}$ (2P) &$\frac{3}{2}^{-}$ (2P)\\ \hline
Mass &6072 & 6072 & 6238 & 6238 &6249 & 6249  \\ \hline
$\Sigma_{b}^{+}$$\pi^{-}$&1.39&5.50$\times10^{-1}$&1.98&11.5&2.40$\times10^{-2}$&2.76 \\    \hline
$\Sigma_{b}^{0}$$\pi^{0}$&1.39&5.40$\times10^{-1}$&1.97&11.5&2.40$\times10^{-2}$&2.73 \\    \hline
$\Sigma_{b}^{-}$$\pi^{+}$&1.23&4.90$\times10^{-1}$&1.94&11.7&2.20$\times10^{-2}$& 2.60\\    \hline
$\Sigma_{b}^{*+}$$\pi^{-}$&1.68&7.00$\times10^{-1}$&2.80$\times10^{-2}$&3.76&1.93&14.0 \\    \hline
$\Sigma_{b}^{*0}$$\pi^{0}$&1.69&7.00$\times10^{-1}$&2.80$\times10^{-2}$&3.71&1.92&14.0 \\    \hline
$\Sigma_{b}^{*-}$$\pi^{+}$&1.44&6.10$\times10^{-1}$&2.60$\times10^{-2}$&3.51&1.89&14.0 \\    \hline
$\Gamma_{total}$&8.82&3.59&5.97&45.68&5.81&50.09 \\
\end{tabular}
\end{ruledtabular}
\end{table*}
\begin{table*}[h]
\begin{ruledtabular}\caption{Two-body strong decays of $\Lambda_{b}$ in MeV}
\label{Table VII}
\begin{tabular}{c c c c c c c c  }
N&1&2&3&4&5&6  \\  \hline
Notations & $\Lambda_{b2}^{2}$&$\Lambda_{b2}^{2\lambda^{\prime} }$&$\Lambda_{b2}^{2\rho^{\prime} }$&$\Lambda_{b2}^{2}$&$\Lambda_{b2}^{2\lambda^{\prime} }$&$\Lambda_{b2}^{2\rho^{\prime} }$ \\ \hline
 Assignments&  $\frac{3}{2}^{+}$ (1D)& $\frac{3}{2}^{+}$ (2D)& $\frac{3}{2}^{+}$ (2D)& $\frac{5}{2}^{+}$ (1D) & $\frac{5}{2}^{+}$ (2D)& $\frac{5}{2}^{+}$ (2D) \\ \hline
Mass &6146 & 6432 & 6432 & 6153 &6440 & 6440  \\ \hline
$\Sigma_{b}^{+}$$\pi^{-}$&1.75&2.95&3.70$\times10^{-1}$&1.30$\times10^{-2}$&3.50$\times10^{-1}$&4.20$\times10^{-1}$  \\    \hline
$\Sigma_{b}^{0}$$\pi^{0}$&1.74&2.94&3.80$\times10^{-1}$&1.30$\times10^{-2}$&3.50$\times10^{-1}$&4.10$\times10^{-1}$ \\    \hline
$\Sigma_{b}^{-}$$\pi^{+}$&1.65&2.91&3.90$\times10^{-1}$&1.10$\times10^{-2}$&3.30$\times10^{-1}$&4.00$\times10^{-1}$ \\    \hline
$\Sigma_{b}^{*+}$$\pi^{-}$&2.90$\times10^{-1}$&1.02&6.70$\times10^{-1}$&1.84&3.67&7.60$\times10^{-1}$\\    \hline
$\Sigma_{b}^{*0}$$\pi^{0}$&2.90$\times10^{-1}$&1.01&6.60$\times10^{-1}$&1.82&3.65&7.60$\times10^{-1}$\\    \hline
$\Sigma_{b}^{*-}$$\pi^{+}$&2.70$\times10^{-1}$&0.99&6.40$\times10^{-1}$&1.73&3.60&7.70$\times10^{-1}$\\    \hline
$\Lambda_{b}$$\omega$&-&4.50$\times10^{-1}$&5.30$\times10^{-1}$&-&6.40$\times10^{-1}$&7.20$\times10^{-1}$ \\    \hline
$B^{*-}$$p$&-&2.72&8.80$\times10^{-2}$&-&2.85&1.10$\times10^{-1}$ \\    \hline
$B^{*0}$$n$&-&2.71&8.70$\times10^{-2}$&-&2.84&1.10$\times10^{-1}$ \\    \hline
$\Gamma_{total}$&5.99&17.7&3.82&5.43&18.28&4.46 \\
\end{tabular}
\end{ruledtabular}
\end{table*}
\begin{table*}[h]
\begin{ruledtabular}\caption{Two-body strong decays of $\Sigma_{c}$ in MeV}
\label{Table VIII}
\begin{tabular}{c c c c c c c c }
N&1&2&3&4&5&6&7 \\  \hline
Notations &$\Sigma_{c1}^{0\lambda^{\prime}}$ &$\Sigma_{c1}^{0\rho^{\prime}}$& $\Sigma_{c1}^{0}$ & $\Sigma_{c1}^{0\lambda^{\prime}}$ &$\Sigma_{c1}^{0\rho^{\prime}}$ &$\Sigma_{c0}^{1}$ & $\Sigma_{c0}^{1\lambda^{\prime}}$ \\ \hline
Assignments& $\frac{1}{2}^{+}$ (2S)& $\frac{1}{2}^{+}$ (2S)& $\frac{3}{2}^{+}$ (1S)&$\frac{3}{2}^{+}$ (2S)& $\frac{3}{2}^{+}$ (2S)& $\frac{1}{2}^{-}$ (1P) & $\frac{1}{2}^{-}$ (2P) \\ \hline
Mass &2913 & 2913 & 2518 & 2967 &2967 & 2823&3196  \\ \hline
$\Lambda_{c}$$\pi^{+}$&71.70&5.38$\times10^{-1}$&12.90&93.70&1.75$\times10^{-4}$&282.70&6.96 \\    \hline
$\Lambda_{c}$$\rho^{+}$&-&-&-&-&-&-&1.26$\times10^{-9}$ \\    \hline
$\Sigma_{c}^{++}$$\pi^{0}$&33.00&3.80&-&12.40&8.26$\times10^{-1}$&1.23$\times10^{-10}$&8.93$\times10^{-10}$ \\    \hline
$\Sigma_{c}^{+}$$\pi^{+}$&32.70&3.81&-&12.30&8.33$\times10^{-1}$&1.13$\times10^{-10}$&3.03$\times10^{-9}$\\    \hline
$\Sigma_{c}^{*++}$$\pi^{0}$&9.28&1.73&-&38.00&4.78&1.08$\times10^{-11}$&3.55$\times10^{-9}$ \\    \hline
$\Sigma_{c}^{*+}$$\pi^{+}$&9.28&1.73&-&38.00&4.78&1.08$\times10^{-11}$&3.55$\times10^{-9}$ \\    \hline
$\Xi_{c}^{'+}$$K^{+}$&-&-&-&-&-&-&1.50$\times10^{-10}$ \\    \hline
$\Xi_{c}^{+}$$K^{+}$&-&-&-&2.74$\times10^{-3}$&1.35$\times10^{-3}$&-&2.31 \\    \hline
$\Xi_{c}^{*'+}$$K^{+}$&-&-&-&-&-&-&1.41$\times10^{-10}$ \\    \hline
$D^{+}$$p$&10.20&3.50$\times10^{-1}$&-&18.80&1.26&100.30&8.09 \\    \hline
$D^{*+}$$p$&-&-&-&1.77&1.71$\times10^{-3}$&-&1.38$\times10^{-8}$\\    \hline
$D^{0}$$\Delta^{++}$&-&-&-&-&-&-&7.91$\times10^{-9}$\\    \hline
$D^{+}$$\Delta^{+}$&-&-&-&-&-&-&2.36$\times10^{-9}$\\    \hline
$D_{s}$$\Sigma^{+}$&-&-&-&-&-&-&1.51$\times10^{-11}$\\    \hline
$\Gamma_{total}$&166.16&11.96&12.90&214.97&12.48&383&17.36 \\
\end{tabular}
\end{ruledtabular}
\end{table*}
\begin{table*}[h]
\begin{ruledtabular}\caption{Two-body strong decays of $\Sigma_{c}$ in MeV}
\label{Table IX}
\begin{tabular}{c c c c c c c c  }
N&1&2&3&4&5&6&7  \\  \hline
Notations &$\Sigma_{c0}^{1\rho^{\prime}}$ &$\Sigma_{c1}^{1}$& $\Sigma_{c1}^{1\lambda^{\prime}}$ & $\Sigma_{c1}^{1\rho^{\prime}}$ &$\Sigma_{c1}^{1}$ &$\Sigma_{c1}^{1\lambda^{\prime}}$ & $\Sigma_{c1}^{1\rho^{\prime}}$ \\ \hline
Assignments& $\frac{1}{2}^{-}$ (2P)& $\frac{1}{2}^{-}$ (1P)& $\frac{1}{2}^{-}$ (2P)& $\frac{1}{2}^{-}$ (2P)& $\frac{3}{2}^{-}$ (1P)& $\frac{3}{2}^{-}$ (2P) & $\frac{3}{2}^{-}$ (2P) \\ \hline
Mass &3196 & 2809 & 3185 & 3185 &2829 & 3202&3202  \\ \hline
$\Lambda_{c}$$\pi^{+}$&16.30&5.85$\times10^{-11}$&4.53$\times10^{-9}$&7.91$\times10^{-10}$&2.22$\times10^{-10}$&4.45$\times10^{-9}$&4.27$\times10^{-10}$ \\    \hline
$\Lambda_{c}$$\rho^{+}$&3.15$\times10^{-9}$&-&3.42&57.8&-&3.69&52.80 \\    \hline
$\Sigma_{c}^{++}$$\pi^{0}$&1.25$\times10^{-10}$&157&5.85&2.90$\times10^{-2}$&9.92$\times10^{-1}$&2.97$\times10^{-1}$&9.50 \\    \hline
$\Sigma_{c}^{+}$$\pi^{+}$&2.36$\times10^{-10}$&157&5.85&3.20$\times10^{-2}$&9.71$\times10^{-1}$&2.95$\times10^{-1}$&9.48 \\    \hline
$\Sigma_{c}^{*++}$$\pi^{0}$&8.65$\times10^{-10}$&3.90$\times10^{-1}$&2.84$\times10^{-1}$&13.20&134&5.99&8.42 \\    \hline
$\Sigma_{c}^{*+}$$\pi^{+}$&8.65$\times10^{-10}$&3.90$\times10^{-1}$&2.84$\times10^{-1}$&13.20&134&5.99&8.42 \\    \hline
$\Xi_{c}^{'+}$$K^{+}$&1.40$\times10^{-11}$&-&1.02&9.56&-&1.36$\times10^{-3}$&2.07$\times10^{-1}$ \\    \hline
$\Xi_{c}^{+}$$K^{+}$&6.68&-&3.82$\times10^{-10}$&7.40$\times10^{-11}$&-&1.54$\times10^{-10}$&1.25$\times10^{-10}$ \\    \hline
$\Xi_{c}^{*'+}$$K^{+}$&3.04$\times10^{-11}$&-&1.07$\times10^{-4}$&2.60$\times10^{-2}$&-&7.15$\times10^{-1}$&9.69 \\    \hline
$D^{+}$$p$&8.23&1.26$\times10^{-12}$&8.45$\times10^{-11}$&4.65$\times10^{-11}$&7.76$\times10^{-12}$&6.24$\times10^{-9}$&1.80$\times10^{-10}$ \\    \hline
$D^{*+}$$p$&7.51$\times10^{-13}$&-&6.33&1.68&-&6.34&2.08\\    \hline
$D^{0}$$\Delta^{++}$&1.88$\times10^{-11}$&-&6.30$\times10^{-2}$&5.38$\times10^{-1}$&-&10.4&6.60$\times10^{-1}$\\    \hline
$D^{+}$$\Delta^{+}$&2.52$\times10^{-11}$&-&1.90$\times10^{-2}$&1.63$\times10^{-1}$&-&3.42&2.50$\times10^{-1}$\\    \hline
$D_{s}$$\Sigma^{+}$&1.40$\times10^{-11}$&-&2.95&1.68&-&1.58$\times10^{-3}$&1.90$\times10^{-2}$\\    \hline
$\Gamma_{total}$&31.21&314.78&26.07&97.91&269.96&37.14&101.53 \\
\end{tabular}
\end{ruledtabular}
\end{table*}
\begin{table*}[h]
\begin{ruledtabular}\caption{Two-body strong decays of $\Sigma_{c}$ in MeV}
\label{Table X}
\begin{tabular}{c c c c c c c c  }
N&1&2&3&4&5&6&~  \\  \hline
Notations &$\Sigma_{c2}^{1}$& $\Sigma_{c2}^{1\lambda^{\prime}}$ & $\Sigma_{c2}^{1\rho^{\prime}}$ &$\Sigma_{c2}^{1}$ &$\Sigma_{c2}^{1\lambda^{\prime}}$ & $\Sigma_{c2}^{1\rho^{\prime}}$ \\ \hline
Assignments& $\frac{3}{2}^{-}$ (1P)& $\frac{3}{2}^{-}$ (2P)& $\frac{3}{2}^{-}$ (2P)& $\frac{5}{2}^{-}$ (1P)& $\frac{5}{2}^{-}$ (2P)& $\frac{5}{2}^{-}$ (2P) &~ \\ \hline
Mass &2806 & 3179 & 3179 & 2835 &3207 & 3207  \\ \hline
$\Lambda_{c}$$\pi^{+}$&13.60&2.06&33.40&18.70&2.49&35.0&~ \\    \hline
$\Lambda_{c}$$\rho^{+}$&-&1.70$\times10^{-2}$&8.99&-&3.70$\times10^{-2}$&14.5&~ \\    \hline
$\Sigma_{c}^{++}$$\pi^{0}$&1.13&4.36$\times10^{-1}$&15.57&8.70$\times10^{-1}$&2.46$\times10^{-1}$&7.75&~ \\    \hline
$\Sigma_{c}^{+}$$\pi^{+}$&1.11&4.33$\times10^{-1}$&5.53&8.50$\times10^{-1}$&2.46$\times10^{-1}$&7.74&~ \\    \hline
$\Sigma_{c}^{*++}$$\pi^{0}$&2.95$\times10^{-1}$&2.41$\times10^{-1}$&11.51&9.70$\times10^{-1}$&4.93$\times10^{-1}$&20.70&~ \\    \hline
$\Sigma_{c}^{*+}$$\pi^{+}$&2.95$\times10^{-1}$&2.41$\times10^{-1}$&11.51&9.70$\times10^{-1}$&4.93$\times10^{-1}$&20.70&~ \\    \hline
$\Xi_{c}^{'+}$$K^{+}$&-&1.29$\times10^{-3}$&2.30$\times10^{-1}$&-&1.24$\times10^{-3}$&1.83$\times10^{-1}$&~ \\    \hline
$\Xi_{c}^{+}$$K^{+}$&-&1.90$\times10^{-2}$&1.75&-&3.00$\times10^{-2}$&2.33&~ \\    \hline
$\Xi_{c}^{*'+}$$K^{+}$&-&6.28$\times10^{-5}$&1.6$\times10^{-2}$&-&4.96$\times10^{-4}$&1.07$\times10^{-1}$&~ \\    \hline
$D^{+}$$p$&-&0.854&1.95&5.70$\times10^{-2}$&1.05&1.99&~ \\    \hline
$D^{*+}$$p$&-&3.60$\times10^{-1}$&2.66&-&4.98$\times10^{-1}$&3.16&~\\    \hline
$D^{0}$$\Delta^{++}$&-&4.80$\times10^{-2}$&4.30$\times10^{-1}$ &-&1.43$\times10^{-1}$&1.08&~\\    \hline
$D^{+}$$\Delta^{+}$&-&1.40$\times10^{-2}$&1.30$\times10^{-1}$ &-&4.30$\times10^{-2}$&3.40$\times10^{-1}$&~\\    \hline
$D_{s}$$\Sigma^{+}$&-&4.55$\times10^{-4}$&6.25$\times10^{-3}$ &-&1.67$\times10^{-3}$&2.00$\times10^{-2}$&~\\    \hline
$\Gamma_{total}$&19.13&4.72&93.68 &22.42&5.77&115.60 \\
\end{tabular}
\end{ruledtabular}
\end{table*}
\begin{table*}[h]
\begin{ruledtabular}\caption{Two-body strong decays of $\Sigma_{b}$ in MeV}
\label{Table XI}
\begin{tabular}{c c c c c c c c }
N&1&2&3&4&5&6&7 \\  \hline
Notations &$\Sigma_{b1}^{0\lambda^{\prime}}$ &$\Sigma_{b1}^{0\rho^{\prime}}$& $\Sigma_{b1}^{0}$ & $\Sigma_{b1}^{0\lambda^{\prime}}$ &$\Sigma_{b1}^{0\rho^{\prime}}$ &$\Sigma_{b0}^{1}$ & $\Sigma_{b0}^{1\lambda^{\prime}}$ \\ \hline
Assignments& $\frac{1}{2}^{+}$ (2S)&$\frac{1}{2}^{+}$ (2S)& $\frac{3}{2}^{+}$ (1S)& $\frac{3}{2}^{+}$ (2S)& $\frac{3}{2}^{+}$ (2S)& $\frac{1}{2}^{-}$ (1P) & $\frac{1}{2}^{-}$ (2P) \\ \hline
Mass &6225 & 6225 & 5835 & 6246 &6246 & 6113 & 6447  \\ \hline
$\Lambda_{b}$$\pi^{+}$&88.80&3.05$\times10^{-1}$&14.0&100&4.50$\times10^{-2}$&316&8.23 \\    \hline
$\Lambda_{b}$$\rho^{+}$&-&-&-&-&-&-&3.42$\times10^{-9}$ \\    \hline
$\Sigma_{b}^{+}$$\pi^{0}$&28.40&4.09&-&8.67&1.04&3.64$\times10^{-12}$&2.60$\times10^{-9}$ \\    \hline
$\Sigma_{b}^{0}$$\pi^{+}$&28.20&4.09&-&8.59&1.04&6.54$\times10^{-12}$&2.56$\times10^{-9}$\\    \hline
$\Sigma_{b}^{*+}$$\pi^{0}$&11.80&1.96&-&36.3&5.13&6.60$\times10^{-13}$&1.62$\times10^{-9}$ \\    \hline
$\Sigma_{b}^{*0}$$\pi^{+}$&11.80&1.96&-&36.3&5.13&6.60$\times10^{-13}$&1.62$\times10^{-9}$ \\    \hline
$\Xi_{b}^{'+}$$K^{+}$&-&-&-&-&-&-&- \\    \hline
$\Xi_{b}^{+}$$K^{+}$&-&-&-&-&-&-&2.31 \\    \hline
$\overline{B}^{0}$$p$&3.06$\times10^{-1}$&3.19$\times10^{-3}$&-&2.46&4.70$\times10^{-2}$&-&12.4 \\    \hline
$\overline{B}^{*0}$$p$&-&-&-&-&-&-&7.11$\times10^{-11}$\\    \hline
$\Gamma_{total}$&169.31&12.41&14.00&189.91&12.43&316.00&22.94 \\
\end{tabular}
\end{ruledtabular}
\end{table*}
\begin{table*}[h]
\begin{ruledtabular}\caption{Two-body strong decays of $\Sigma_{b}$ in MeV}
\label{Table XII}
\begin{tabular}{c c c c c c c c  }
N&1&2&3&4&5&6&7  \\  \hline
Notations &$\Sigma_{b0}^{1\rho^{\prime}}$ &$\Sigma_{b1}^{1}$& $\Sigma_{b1}^{1\lambda^{\prime}}$ & $\Sigma_{b1}^{1\rho^{\prime}}$ &$\Sigma_{b1}^{1}$ &$\Sigma_{b1}^{1\lambda^{\prime}}$ & $\Sigma_{b1}^{1\rho^{\prime}}$ \\ \hline
Assignments& $\frac{1}{2}^{-}$ (2P)& $\frac{1}{2}^{-}$ (1P)& $\frac{1}{2}^{-}$ (2P)& $\frac{1}{2}^{-}$ (2P)& $\frac{3}{2}^{-}$ (1P)& $\frac{3}{2}^{-}$ (2P) & $\frac{3}{2}^{-}$ (2P) \\ \hline
Mass &6447 & 6107 & 6442 & 6442 &6116 & 6450 & 6450  \\ \hline
$\Lambda_{b}$$\pi^{+}$&16.70&9.39$\times10^{-11}$&5.50$\times10^{-9}$&1.06$\times10^{-9}$&-&6.05$\times10^{-9}$&1.37$\times10^{-9}$ \\    \hline
$\Lambda_{b}$$\rho^{+}$&1.87$\times10^{-10}$&-&2.47&83.4&-&2.71&83.3 \\    \hline
$\Sigma_{b}^{+}$$\pi^{0}$&4.47$\times10^{-10}$&139&6.45&2.08&3.52$\times10^{-1}$&1.80$\times10^{-1}$&7.89 \\    \hline
$\Sigma_{b}^{0}$$\pi^{+}$&4.82$\times10^{-10}$&138&6.45&2.14&3.40$\times10^{-1}$&1.79$\times10^{-1}$&7.86 \\    \hline
$\Sigma_{b}^{*+}$$\pi^{0}$&6.71$\times10^{-10}$&3.59$\times10^{-1}$&2.66$\times10^{-1}$&13.4&132&6.53&9.90 \\    \hline
$\Sigma_{b}^{*0}$$\pi^{+}$&6.71$\times10^{-10}$&3.59$\times10^{-1}$&2.66$\times10^{-1}$&13.4&132&6.53&9.90 \\    \hline
$\Xi_{b}^{'+}$$K^{+}$&1.40$\times10^{-12}$&-&9.70$\times10^{-2}$&1.92&-&1.43$\times10^{-6}$&3.77$\times10^{-4}$  \\    \hline
$\Xi_{b}^{+}$$K^{+}$&12.40&-&1.54$\times10^{-10}$&1.84$\times10^{-11}$&-&1.87$\times10^{-7}$&2.56$\times10^{-11}$ \\    \hline
$\overline{B}^{0}$$p$&4.60&-&2.62$\times10^{-10}$&1.51$\times10^{-10}$&-&4.69$\times10^{-10}$&5.82$\times10^{-11}$ \\    \hline
$\overline{B}^{*0}$$p$&7.65$\times10^{-11}$&-&8.39&1.67&-&8.41&1.96\\    \hline
$\Gamma_{total}$&33.70&277.72&24.39&118.01&264.69&24.54&120.81 \\
\end{tabular}
\end{ruledtabular}
\end{table*}
\begin{table*}[h]
\begin{ruledtabular}\caption{Two-body strong decays of $\Sigma_{b}$ in MeV}
\label{Table XIII}
\begin{tabular}{c c c c c c c c }
N&1&2&3&4&5&6  \\  \hline
Notations&$\Sigma_{b2}^{1}$& $\Sigma_{b2}^{1\lambda^{\prime}}$ & $\Sigma_{b2}^{1\rho^{\prime}}$ &$\Sigma_{b2}^{1}$ &$\Sigma_{b2}^{1\lambda^{\prime}}$ & $\Sigma_{b2}^{1\rho^{\prime}}$ \\ \hline
Assignments& $\frac{3}{2}^{-}$ (1P)& $\frac{3}{2}^{-}$ (2P)& $\frac{3}{2}^{-}$ (2P)& $\frac{5}{2}^{-}$ (1P)& $\frac{5}{2}^{-}$ (2P)&$\frac{5}{2}^{-}$ (2P) \\ \hline
Mass &6098 & 6439 & 6439 & 6119 &6452 & 6452   \\ \hline
$\Lambda_{b}$$\pi^{+}$&14.10&2.34&38.6&16.7&2.60&39.7\\    \hline
$\Lambda_{b}$$\rho^{+}$&-&1.15$\times10^{-3}$ &1.53&-&2.62$\times10^{-3}$&2.86\\    \hline
$\Sigma_{b}^{+}$$\pi^{0}$&4.81$\times10^{-1}$&2.87$\times10^{-1}$&13.30&3.00$\times10^{-1}$&1.47$\times10^{-1}$&6.39 \\    \hline
$\Sigma_{b}^{0}$$\pi^{+}$&4.62$\times10^{-1}$&2.84$\times10^{-1}$&13.20&2.91$\times10^{-1}$&1.46$\times10^{-1}$&6.36 \\    \hline
$\Sigma_{b}^{*+}$$\pi^{0}$&2.99$\times10^{-1}$&2.31$\times10^{-1}$&11.8&6.80$\times10^{-1}$&4.19$\times10^{-1}$&20.00 \\    \hline
$\Sigma_{b}^{*0}$$\pi^{+}$&2.99$\times10^{-1}$&2.31$\times10^{-1}$&11.8&6.80$\times10^{-1}$&4.19$\times10^{-1}$&20.00\\    \hline
$\Xi_{b}^{'+}$$K^{+}$&-&- &-&-&8.88$\times10^{-7}$&5.02$\times10^{-4}$ \\    \hline
$\Xi_{b}^{+}$$K^{+}$&-&8.11$\times10^{-3}$&1.00&-&1.10$\times10^{-2}$&1.24 \\    \hline
$\overline{B}^{0}$$p$&-&5.13$\times10^{-1}$&1.51&-&6.03$\times10^{-1}$&1.61\\    \hline
$\overline{B}^{*0}$$p$&-&4.09$\times10^{-1}$&2.42&-&5.00$\times10^{-1}$&2.71\\    \hline
$\Gamma_{total}$&15.64&4.30&95.16&18.65&4.85&100.87\\
\end{tabular}
\end{ruledtabular}
\end{table*}
\begin{table*}[h]
\begin{ruledtabular}\caption{Two-body strong decays of $\Omega_{c}$ in MeV}
\label{Table XIV}
\begin{tabular}{c c c c c c c c }
N&1&2&3&4&5&6 \\  \hline
Notations&$\Omega_{c1}^{0\lambda^{\prime}}$& $\Omega_{c1}^{0\rho^{\prime}}$ &$\Omega_{c1}^{0\lambda^{\prime}}$ &$\Omega_{c1}^{0\rho^{\prime}}$ & $\Omega_{c0}^{1}$ & $\Omega_{c0}^{1\lambda^{\prime}}$ \\ \hline
Assignments& $\frac{1}{2}^{+}$ (2S)& $\frac{1}{2}^{+}$ (2S)& $\frac{3}{2}^{+}$ (2S)& $\frac{3}{2}^{+}$ (2S)& $\frac{1}{2}^{-}$ (1P) & $\frac{1}{2}^{-}$ (2P) \\ \hline
Mass &3120 &3120  & 3197 & 3197 & 3057 & 3426  \\ \hline
$\Xi_{c}^{+}$$K^{-}$&14.40&2.03&28.9&1.38&183&4.79 \\    \hline
$\Xi_{c}^{+}$$K^{*-}$&-&-&-&-&-&1.11$\times10^{-8}$ \\    \hline
$\Xi_{c}^{0}$$\overline{K}^{0}$&13.50&2.04&27.81&1.47&181&4.83 \\    \hline
$\Xi_{c}^{0}$$\overline{K}^{*0}$&-&-&-&-&-&3.54$\times10^{-9}$\\    \hline
$\Xi_{c}^{'+}$$K^{-}$&2.77&1.01&3.28&6.40$\times10^{-1}$&-&7.45$\times10^{-10}$ \\    \hline
$\Xi_{c}^{*'+}$$K^{-}$&-&-&3.17&1.18&-&1.15$\times10^{-10}$ \\    \hline
$\Xi_{c}^{'0}$$\overline{K}^{0}$&2.29&8.70$\times10^{-1}$&3.07&6.30$\times10^{-1}$&-&1.51$\times10^{-9}$ \\    \hline
$\Xi_{c}^{*'0}$$\overline{K}^{0}$&-&-&2.69&1.03&-&5.34$\times10^{-10}$ \\    \hline
$\Xi^{0}$$D^{0}$&-&-&5.60$\times10^{-1}$&2.31$\times10^{-3}$ &-&9.90$\times10^{-10}$ \\    \hline
$\Xi^{-}$$D^{+}$&-&-&9.00$\times10^{-2}$&1.67$\times10^{-4}$ &-&3.80$\times10^{-10}$\\    \hline
$\Xi^{0}$$D^{*0}$&-&-&-&-&-&1.85 \\    \hline
$\Xi^{-}$$D^{*+}$&-&-&-&-&-&1.77\\    \hline
$\Gamma_{total}$&32.96&5.95&78.16&6.33&364.00&13.24 \\
\end{tabular}
\end{ruledtabular}
\end{table*}
\begin{table*}[h]
\begin{ruledtabular}\caption{Two-body strong decays of $\Omega_{c}$ in MeV}
\label{Table XV}
\begin{tabular}{c c c c c c c c }
N&1&2&3&4&5&6&7 \\  \hline
Notations&$\Omega_{c0}^{1\rho^{\prime}}$& $\Omega_{c1}^{1}$ & $\Omega_{c1}^{1\lambda^{\prime}}$ &$\Omega_{c1}^{1\rho^{\prime}}$ &$\Omega_{c1}^{1}$ & $\Omega_{c1}^{1\lambda^{\prime}}$ & $\Omega_{c1}^{1\rho^{\prime}}$ \\ \hline
Assignments& $\frac{1}{2}^{-}$ (2P)& $\frac{1}{2}^{-}$ (1P)& $\frac{1}{2}^{-}$ (2P) &$\frac{1}{2}^{-}$ (2P)& $\frac{3}{2}^{-}$ (1P)& $\frac{3}{2}^{-}$ (2P) & $\frac{3}{2}^{-}$ (2P) \\ \hline
Mass &3426 &3000 & 3416 & 3416 & 3050 & 3431 & 3431  \\ \hline
$\Xi_{c}^{+}$$K^{-}$&2.25&1.30$\times10^{-11}$&1.38$\times10^{-9}$&2.89$\times10^{-9}$&2.06$\times10^{-10}$&7.89$\times10^{-10}$&2.27$\times10^{-9}$ \\    \hline
$\Xi_{c}^{+}$$K^{*-}$&7.90$\times10^{-10}$&-&1.21&39.60&-&1.41&38.9 \\    \hline
$\Xi_{c}^{0}$$\overline{K}^{0}$&1.95&8.76$\times10^{-12}$&1.25$\times10^{-9}$&9.24$\times10^{-11}$&7.88$\times10^{-11}$&4.80$\times10^{-10}$&2.49$\times10^{-9}$ \\    \hline
$\Xi_{c}^{0}$$\overline{K}^{*0}$&1.60$\times10^{-9}$&-&1.18&39.6&-&1.38&39.10\\    \hline
$\Xi_{c}^{'+}$$K^{-}$&9.67$\times10^{-11}$&-&3.52&8.94$\times10^{-1}$&-&1.14$\times10^{-1}$&4.61 \\    \hline
$\Xi_{c}^{*'+}$$K^{-}$&5.67$\times10^{-10}$&-&7.25$\times10^{-2}$&4.95&-&3.44&7.30 \\    \hline
$\Xi_{c}^{'0}$$\overline{K}^{0}$&3.15$\times10^{-10}$&-&3.53&1.05&-&1.09$\times10^{-1}$&4.53 \\    \hline
$\Xi_{c}^{*'0}$$\overline{K}^{0}$&4.88$\times10^{-10}$&-&6.94$\times10^{-2}$&4.83&-&3.43&7.51 \\    \hline
$\Xi^{0}$$D^{0}$&9.84$\times10^{-12}$&-&11.0&2.602&-&3.40$\times10^{-1}$&1.19 \\    \hline
$\Xi^{-}$$D^{+}$&5.07$\times10^{-11}$&-&11.1&2.012&-&3.02$\times10^{-1}$&1.14\\    \hline
$\Xi^{0}$$D^{*0}$&2.69&-&1.56$\times10^{-1}$&2.097&-&1.16$\times10^{-1}$&1.42 \\    \hline
$\Xi^{-}$$D^{*+}$&2.38&-&1.18$\times10^{-1}$&1.67&-&9.10$\times10^{-2}$&1.18\\    \hline
$\Gamma_{total}$&9.27&2.18$\times10^{-11}$&31.96&99.31&2.85$\times10^{-10}$&10.73&106.88 \\
\end{tabular}
\end{ruledtabular}
\end{table*}
\begin{table*}[h]
\begin{ruledtabular}\caption{Two-body strong decays of $\Omega_{c}$ in MeV}
\label{Table XVI}
\begin{tabular}{c c c c c c c c }
N&1&2&3&4&5&6&~ \\  \hline
Notations&$\Omega_{c2}^{1}$& $\Omega_{c2}^{1\lambda^{\prime}}$ & $\Omega_{c2}^{1\rho^{\prime}}$ &$\Omega_{c2}^{1}$  & $\Omega_{c2}^{1\lambda^{\prime}}$ & $\Omega_{c2}^{1\rho^{\prime}}$ &~ \\ \hline
Assignments& $\frac{3}{2}^{-}$ (1P)& $\frac{3}{2}^{-}$ (2P)& $\frac{3}{2}^{-}$ (2P)& $\frac{5}{2}^{-}$ (1P)& $\frac{5}{2}^{-}$ (2P)& $\frac{5}{2}^{-}$ (2P) &~ \\ \hline
Mass &3066 &3411 & 3411 & 3090 & 3435 & 3435 & ~  \\ \hline
$\Xi_{c}^{+}$$K^{-}$&5.99$\times10^{-1}$&0.634&15.80&1.35&7.80$\times10^{-1}$&17.0 \\    \hline
$\Xi_{c}^{+}$$K^{*-}$&-&5.81$\times10^{-4}$&7.68$\times10^{-1}$&-&2.29$\times10^{-3}$&2.04 \\    \hline
$\Xi_{c}^{0}$$\overline{K}^{0}$&4.80$\times10^{-1}$&6.10$\times10^{-1}$&15.6&1.15&7.51$\times10^{-1}$&16.8 \\    \hline
$\Xi_{c}^{0}$$\overline{K}^{*0}$&-&5.07$\times10^{-4}$&6.92$\times10^{-1}$&-&2.15$\times10^{-3}$&1.91\\    \hline
$\Xi_{c}^{'+}$$K^{-}$&-&1.64$\times10^{-1}$&7.43&-&9.50$\times10^{-2}$&3.77 \\    \hline
$\Xi_{c}^{*'+}$$K^{-}$&-&6.10$\times10^{-2}$&4.27&-&1.32$\times10^{-1}$&8.08 \\    \hline
$\Xi_{c}^{'0}$$\overline{K}^{0}$&-&1.57$\times10^{-1}$&7.27&-&9.10$\times10^{-2}$&3.70 \\    \hline
$\Xi_{c}^{*'0}$$\overline{K}^{0}$&-&5.80$\times10^{-2}$&4.16&-&1.26$\times10^{-1}$&7.90 \\    \hline
$\Xi^{0}$$D^{0}$&-&4.88$\times10^{-1}$&1.95&-&2.84$\times10^{-1}$&9.67$\times10^{-1}$ \\    \hline
$\Xi^{-}$$D^{+}$&-&4.28$\times10^{-1}$&1.84&-&2.52$\times10^{-1}$&9.26$\times10^{-1}$\\    \hline
$\Xi^{0}$$D^{*0}$&-&16.0&6.67&-&5.10$\times10^{-2}$&6.12$\times10^{-1}$ \\    \hline
$\Xi^{-}$$D^{*+}$&-&15.4&7.62&-&4.00$\times10^{-2}$&5.12$\times10^{-1}$\\    \hline
$\Gamma_{total}$&1.08&34.00&74.07&2.50&2.61&64.22 \\
\end{tabular}
\end{ruledtabular}
\end{table*}
\begin{table*}[h]
\begin{ruledtabular}\caption{Two-body strong decays of $\Omega_{c}$ in MeV}
\label{Table XVII}
\begin{tabular}{c c c c c c c c }
N&1&2&3&4&5&6&~ \\  \hline
Notations&$\Omega_{c1}^{2}$& $\Omega_{c1}^{2}$ & $\Omega_{c2}^{2}$ &$\Omega_{c2}^{2}$  & $\Omega_{c3}^{2}$ & $\Omega_{c3}^{2}$ &~ \\ \hline
Assignments& $\frac{1}{2}^{+}$ (1D)& $\frac{3}{2}^{+}$ (1D)& $\frac{3}{2}^{+}$ (1D)& $\frac{5}{2}^{+}$ (1D)& $\frac{5}{2}^{+}$ (1D)& $\frac{7}{2}^{+}$ (1D) &~ \\ \hline
Mass &3304 &3313 & 3304 & 3314 & 3304 & 3315 & ~  \\ \hline
$\Xi_{c}^{+}$$K^{-}$             & 10.01 & 10.12 & - & - & 1.04 &1.17 \\    \hline
$\Xi_{c}^{0}$$\overline{K}^{0}$  & 9.98  & 10.97 & - & - & 9.82$\times10^{-1}$ &1.11 \\    \hline
$\Xi_{c}^{'+}$$K^{-}$            & 2.55  & 6.61$\times10^{-1}$ & 5.74 & 1.54$\times10^{-1}$ & 1.51$\times10^{-1}$ &1.11$\times10^{-1}$ \\    \hline
$\Xi_{c}^{*'+}$$K^{-}$           & 8.00$\times10^{-1}$ & 2.15 & 7.69$\times10^{-1}$ & 4.70 & 4.00$\times10^{-2}$ &7.01$\times10^{-2}$ \\    \hline
$\Xi_{c}^{'0}$$\overline{K}^{0}$ & 2.51 & 6.52$\times10^{-1}$ & 5.65 & 1.44$\times10^{-1}$ & 1.41$\times10^{-1}$ &9.43$\times10^{-2}$ \\    \hline
$\Xi_{c}^{*'0}$$\overline{K}^{0}$& 7.77$\times10^{-1}$ & 2.09 & 7.45$\times10^{-1}$ & 4.58 & 3.64$\times10^{-2}$ &6.42$\times10^{-2}$ \\    \hline
$\Xi^{0}$$D^{0}$                 & 2.23 & 5.97$\times10^{-1}$ & 5.02 & 5.53$\times10^{-2}$ & 4.90$\times10^{-2}$ &- \\ \hline
$\Xi^{-}$$D^{+}$                 & 2.02 & 5.48$\times10^{-1}$ & 4.56 & 4.12$\times10^{-2}$ & 3.54$\times10^{-2}$ &- \\ \hline
$\Gamma_{total}$&30.87&27.73&22.48&9.64&2.48&2.62 \\
\end{tabular}
\end{ruledtabular}
\end{table*}
\begin{table*}[h]
\begin{ruledtabular}\caption{Two-body strong decays of $\Omega_{b}$ in MeV}
\label{Table XVIII}
\begin{tabular}{c c c c c c c c }
N&1&2&3&4&5&6 \\  \hline
Notations&$\Omega_{b1}^{0\lambda^{\prime}}$& $\Omega_{b1}^{0\rho^{\prime}}$ &$\Omega_{b1}^{0\lambda^{\prime}}$ &$\Omega_{b1}^{0\rho^{\prime}}$ & $\Omega_{b0}^{1}$ & $\Omega_{b0}^{1\lambda^{\prime}}$ \\ \hline
Assignments& $\frac{1}{2}^{+}$ (2S)& $\frac{1}{2}^{+}$ (2S)&  $\frac{3}{2}^{+}$ (2S)& $\frac{3}{2}^{+}$ (2S)& $\frac{1}{2}^{-}$ (1P) & $\frac{1}{2}^{-}$ (2P) \\ \hline
Mass &6446 &6446 & 6466 & 6466 & 6334 & 6662  \\ \hline
$\Xi_{b}^{0}$$K^{-}$&258&2.23&25.30&2.03&169&5.83 \\    \hline
$\Xi_{b}^{-}$$\overline{K}^{0}$&235&2.29&23.00&2.16&155&5.91 \\    \hline
$\Xi_{b}^{'0}$$K^{-}$&11.30&3.30$\times10^{-1}$&6.00$\times10^{-1}$&2.30$\times10^{-1}$&-&2.21$\times10^{-10}$ \\    \hline
$\Xi_{b}^{*'0}$$K^{-}$&-&-&3.30$\times10^{-2}$&1.60$\times10^{-2}$&-&8.14$\times10^{-10}$ \\    \hline
$\Xi_{b}^{'-}$$\overline{K}^{0}$&7.03&2.10$\times10^{-1}$&4.87$\times10^{-1}$&1.90$\times10^{-1}$&-&9.17$\times10^{-11}$ \\    \hline
$\Xi_{b}^{*'-}$$\overline{K}^{0}$&-&-&-&-&-&1.19$\times10^{-9}$ \\    \hline
$\Xi^{0}$$B^{-}$&-&-&-&0 &-&9.04$\times10^{-11}$ \\    \hline
$\Xi^{-}$$\overline{B}^{0}$&-&-&-&0 &-&4.98$\times10^{-9}$\\    \hline
$\Xi^{0}$$B^{*-}$&-&-&-&-&-&1.45 \\    \hline
$\Xi^{-}$$\overline{B}^{*0}$&-&-&-&-&-&1.22\\    \hline
$\Gamma_{total}$&511.30&5.06&49.42&4.63&324&14.41 \\
\end{tabular}
\end{ruledtabular}
\end{table*}
\begin{table*}[h]
\begin{ruledtabular}\caption{Two-body strong decays of $\Omega_{b}$ in MeV}
\label{Table XIX}
\begin{tabular}{c c c c c c c c }
N&1&2&3&4&5&6&7 \\  \hline
Notations&$\Omega_{b0}^{1\rho^{\prime}}$& $\Omega_{b1}^{1}$ & $\Omega_{b1}^{1\lambda^{\prime}}$ &$\Omega_{b1}^{1\rho^{\prime}}$ &$\Omega_{b1}^{1}$ & $\Omega_{b1}^{1\lambda^{\prime}}$ & $\Omega_{b1}^{1\rho^{\prime}}$ \\ \hline
Assignments& $\frac{1}{2}^{-}$ (2P)& $\frac{1}{2}^{-}$ (1P)& $\frac{1}{2}^{-}$ (2P) &$\frac{1}{2}^{-}$ (2P)& $\frac{3}{2}^{-}$ (1P)& $\frac{3}{2}^{-}$ (2P) & $\frac{3}{2}^{-}$ (2P) \\ \hline
Mass &6662 &6316 & 6658 & 6658 & 6330 & 6664 & 6664  \\ \hline
$\Xi_{b}^{0}$$K^{-}$&5.60$\times10^{-1}$&2.76$\times10^{-11}$ &6.27$\times10^{-8}$ &8.94$\times10^{-11}$ &6.08$\times10^{-12}$ &1.43$\times10^{-9}$ &1.27$\times10^{-10}$  \\    \hline
$\Xi_{b}^{-}$$\overline{K}^{0}$&2.50$\times10^{-1}$&1.13$\times10^{-11}$ &1.55$\times10^{-8}$ &1.78$\times10^{-10}$ &2.35$\times10^{-11}$ &2.83$\times10^{-9}$ &1.69$\times10^{-10}$  \\    \hline
$\Xi_{b}^{'0}$$K^{-}$&1.68$\times10^{-10}$ &-&3.69&7.33&-&3.90$\times10^{-2}$&2.73 \\    \hline
$\Xi_{b}^{*'0}$$K^{-}$&1.56$\times10^{-10}$ &-&3.80$\times10^{-2}$&3.44&-&3.47&12.70 \\    \hline
$\Xi_{b}^{'-}$$\overline{K}^{0}$&2.04$\times10^{-10}$ &-&3.68&7.76&-&3.70$\times10^{-2}$&2.65 \\    \hline
$\Xi_{b}^{*'-}$$\overline{K}^{0}$&2.25$\times10^{-10}$ &-&3.60$\times10^{-2}$&3.30&-&3.46&13.10 \\    \hline
$\Xi^{0}$$B^{-}$&1.18$\times10^{-11}$ &-&13.90&7.70$\times10^{-1}$&- &3.50$\times10^{-2}$&2.50$\times10^{-1}$ \\    \hline
$\Xi^{-}$$\overline{B}^{0}$&3.01$\times10^{-12}$ &-&13.30&1.16&- &2.60$\times10^{-2}$&1.90$\times10^{-1}$\\    \hline
$\Xi^{0}$$B^{*-}$&1.11&-&6.25$\times10^{-3}$ &9.50$\times10^{-2}$&-&6.51$\times10^{-3}$ &9.50$\times10^{-2}$ \\    \hline
$\Xi^{-}$$\overline{B}^{*0}$&1.01&-&1.82$\times10^{-3}$ &2.90$\times10^{-2}$&-&2.75$\times10^{-3}$ &4.80$\times10^{-2}$\\    \hline
$\Gamma_{total}$&2.93&3.89$\times10^{-11}$&34.65&23.88&2.96$\times10^{-11}$&7.08&31.76 \\
\end{tabular}
\end{ruledtabular}
\end{table*}
\begin{table*}[h]
\begin{ruledtabular}\caption{Two-body strong decays of $\Omega_{b}$ in MeV}
\label{Table XX}
\begin{tabular}{c c c c c c c c }
N&1&2&3&4&5&6&~ \\  \hline
Notations &$\Omega_{b2}^{1}$& $\Omega_{b2}^{1\lambda^{\prime}}$ & $\Omega_{b2}^{1\rho^{\prime}}$ &$\Omega_{b2}^{1}$  & $\Omega_{b2}^{1\lambda^{\prime}}$ & $\Omega_{b2}^{1\rho^{\prime}}$ &~ \\ \hline
Assignments& $\frac{3}{2}^{-}$ (1P)& $\frac{3}{2}^{-}$ (2P)& $\frac{3}{2}^{-}$ (2P)& $\frac{5}{2}^{-}$ (1P)& $\frac{5}{2}^{-}$ (2P)& $\frac{5}{2}^{-}$ (2P) &~ \\ \hline
Mass &6340 &6655 & 6655 & 6350 & 6666 & 6666 & ~  \\ \hline
$\Xi_{b}^{0}$$K^{-}$&1.70$\times10^{-1}$&5.60$\times10^{-1}$&16.50&3..50$\times10^{-1}$&6.30$\times10^{-1}$&17.30&~ \\    \hline
$\Xi_{b}^{-}$$\overline{K}^{0}$&8.10$\times10^{-2}$&5.10$\times10^{-1}$&15.90&2.00$\times10^{-1}$&5.70$\times10^{-1}$&16.70&~ \\    \hline
$\Xi_{b}^{'0}$$K^{-}$&-&6.10$\times10^{-2}$&4.49&-&3.30$\times10^{-2}$&2.23&~ \\    \hline
$\Xi_{b}^{*'0}$$K^{-}$&-&3.20$\times10^{-2}$&2.98&-&6.20$\times10^{-2}$&5.31&~ \\    \hline
$\Xi_{b}^{'-}$$\overline{K}^{0}$&-&5.80$\times10^{-2}$&4.35&-&3.10$\times10^{-2}$&2.16&~ \\    \hline
$\Xi_{b}^{*'-}$$\overline{K}^{0}$&-&3.00$\times10^{-2}$&2.86&-&5.80$\times10^{-2}$&5.11&~ \\    \hline
$\Xi^{0}$$B^{-}$&-&4.40$\times10^{-2}$&3.30$\times10^{-1}$&-&3.00$\times10^{-2}$ &2.10$\times10^{-1}$&~ \\    \hline
$\Xi^{-}$$\overline{B}^{0}$&-&3.10$\times10^{-2}$&2.0$\times10^{-1}$&-&2.30$\times10^{-2}$ &1.70$\times10^{-1}$&~\\    \hline
$\Xi^{0}$$B^{*-}$&-&12.10&9.58&-&3.19$\times10^{-3}$ &4.60$\times10^{-2}$&~\\    \hline
$\Xi^{-}$$\overline{B}^{*0}$&-&9.00&8.23&-&1.46$\times10^{-3}$ &2.20$\times10^{-2}$&~\\    \hline
$\Gamma_{total}$&2.50$\times10^{-1}$&22.43&65.47&5.50$\times10^{-1}$&1.44&49.26&~  \\
\end{tabular}
\end{ruledtabular}
\end{table*}
\begin{table*}[h]
\begin{ruledtabular}\caption{Two-body strong decays of $\Omega_{b}$ in MeV}
\label{Table XXI}
\begin{tabular}{c c c c c c c c }
N&1&2&3&4&5&6&~ \\  \hline
Notations &$\Omega_{b1}^{2}$& $\Omega_{b1}^{2}$ & $\Omega_{b2}^{2}$ &$\Omega_{b2}^{2}$  & $\Omega_{b3}^{2}$ & $\Omega_{b3}^{2}$ &~ \\ \hline
Assignments& $\frac{1}{2}^{+}$ (1D)& $\frac{3}{2}^{+}$ (1D)& $\frac{3}{2}^{+}$ (1D)& $\frac{5}{2}^{+}$ (1D)& $\frac{5}{2}^{+}$ (1D)& $\frac{7}{2}^{+}$ (1D) &~ \\ \hline
Mass &6556 &6561 & 6556 & 6561 & 6556 & 6562 & ~  \\ \hline
$\Xi_{b}^{0}$$K^{-}$&10.82&10.95&-&-&7.83$\times10^{-1}$& 8.51$\times10^{-1}$\\    \hline
$\Xi_{b}^{-}$$\overline{K}^{0}$&10.95&10.74&-&-&6.88$\times10^{-1}$& 7.51$\times10^{-1}$\\    \hline
$\Xi_{b}^{'0}$$K^{-}$&1.61&4.23$\times10^{-1}$&3.62&2.68$\times10^{-2}$&2.65$\times10^{-2}$&1.77$\times10^{-2}$ \\     \hline
$\Xi_{b}^{*'0}$$K^{-}$&5.12$\times10^{-1}$&1.38&4.73$\times10^{-1}$&2.99&-&1.64$\times10^{-2}$ \\     \hline
$\Xi_{b}^{'-}$$\overline{K}^{0}$&1.55&4.08$\times10^{-1}$&3.49&2.40$\times10^{-2}$&2.36$\times10^{-2}$&1.58$\times10^{-2}$ \\     \hline
$\Xi_{b}^{*'-}$$\overline{K}^{0}$&4.82$\times10^{-1}$&1.30&4.44$\times10^{-1}$&2.82&-&1.40$\times10^{-2}$ \\      \hline
$\Gamma_{total}$&25.57&25.20&8.03&5.86&1.53& 1.67 \\
\end{tabular}
\end{ruledtabular}
\end{table*}
\end{document}